\documentclass[aps,prb,twocolumn,showpacs,amsfonts]{revtex4}
\usepackage[dvips]{graphicx,color}

\usepackage{shortvrb}
\MakeShortVerb{\!}
\newcommand{\La} {\line (1,0  ){12}}
\newcommand{\Lb} {\line (3,5  ){6} }
\newcommand{\Lc} {\line (-3,5 ){6} }
\newcommand{\Ld} {\line (-1,0 ){12} }
\newcommand{\Le} {\line (-3,-5){6} }
\newcommand{\Lf} {\line (3,-5) {6} }
\newcommand{\C} {\circle*{5}     }

\newcommand{\pA}{\put(-6,-10)}
\newcommand{\pB}{\put(6,-10)}
\newcommand{\pC}{\put(12,0)}
\newcommand{\pD}{\put(6,10)}
\newcommand{\pE}{\put(-6,10)}
\newcommand{\pF}{\put(-12,0)}

\newcommand{\pG}{\put( 18,-10)}
\newcommand{\pH}{\put( 18,10)}
\newcommand{\pI}{\put(0,20)}
\newcommand{\pJ}{\put(-18,10)}
\newcommand{\pK}{\put(-18,-10)}
\newcommand{\pL}{\put(0,-20)}

\newcommand{\KagHex}{\pA{\C}\pB{\C}\pC{\C}\pD{\C}\pE{\C}\pF{\C}}
\newcommand{\KagStar}{\KagHex\pG{\C}\pH{\C}\pI{\C}\pJ{\C}\pK{\C}\pL{\C}}

\newcommand{\mut}{\tilde{\mu}}
\begin{document}

\title{Quantum dimer model with extensive ground-state entropy on the kagome lattice}

\author{G.~Misguich}
\email{gmisguich@cea.fr}

\author{D.~Serban}
\email{serban@spht.saclay.cea.fr}

\author{V.~Pasquier}
\email{vpasquier@cea.fr}
\affiliation{Service de Physique Th{\'e}orique,
CEA Saclay, 91191 Gif-sur-Yvette Cedex, FRANCE}

\bibliographystyle{prsty}


\begin{abstract}

We introduce a quantum dimer model on  the kagome lattice with kinetic
terms allowing from 3 to 6  dimers to resonate around hexagons. Unlike
models studied previously,  the different resonance loops appears with
different {\em signs} (given  by the  parity  of the number of  dimers
involved).   These   signs   naturally   appear when   performing  the
lowest-order overlap expansion (Rokhsar    and Kivelson 1988)  of  the
Heisenberg model.  We demonstrate that the  quantum dimer model has an
extensive     zero-temperature  entropy and     has   very short-range
dimer-dimer  correlations.  We discuss the  possible relevance of this
new  kind of     quantum dimer    liquid  to  the   physics of     the
spin-$\frac{1}{2}$ Heisenberg model on the kagome lattice.
\end{abstract}

\maketitle



\section{Introduction}

Quantum frustrated Heisenberg antiferromagnets are fascinating systems
which    can   display   a   vast  variety   of    exotic  phases  and
phenomena. Systems with  strong quantum fluctuations where no magnetic
long-range  order develops down to  zero temperature (``spin liquids''
loosely speaking) are of particular interest because  they do not have
direct classical analogs and  are strongly interacting  problems which
resist  to   many  simple     theoretical approaches.     Focusing  on
two-dimensions       and    spin-$\frac{1}{2}$,      two kinds      of
magnetically-disordered   phases  are  well   understood: valence-bond
crystal    (VBC)   and   short-range resonating   valence-bond   (RVB)
liquids. Both are characterized by short-ranged spin-spin correlations
but the VBC has long-ranged   singlet-singlet correlations and  gapped
spin   one   excitations   whereas    RVB  liquid    has  short-ranged
singlet-singlet correlations, topological order and spin-$\frac{1}{2}$
(spinon) excitations.

Despite        of                intense                   theoretical
efforts~\cite{elser89,ze90,mz91,sachdev92,ce92,sh92,ez93,le93,ey94,ze95,nm95,tr96,lblps97,web98,m98,bcl02},
the  physics of  the  spin-$\frac{1}{2}$ kagome\footnote{Kagome comes
from the Japanese words ``kago'' (basket) and  ``me'' (eye).  Although
it is  often written with a capital  K and an   accent on the e  in the
literature, the appropriate spelling  for  a Japanese common noun  is
{\em kagome}.}   antiferromagnetic Heisenberg   model (KAFH) is  still
debated.  For instance, there is still no  consensus on the mechanisms
which produces the unusually large density of  singlet states that was
observed numerically.~\cite{lblps97,web98}

Quantum dimer models~\cite{rk88,msf02}  (QDM) are effective approaches
to  the phases of antiferromagnets  which are dominated by short-range
valence-bonds.  These   models are defined  in  the  Hilbert  space of
nearest-neighbor valence-bond (or dimer)  coverings of the lattice and
contain    kinetic as   well as potential    energy   terms for  these
dimers. These models can often be simpler than  their spin parents and
are   amenable to several analytic   treatments because of their close
relations to  classical dimer  problems~\cite{k63},  Ising models  and
$\mathbb{Z}_2$  gauge theory~\cite{msf02,msp02,ms02}. These models can
offer  simple     descriptions  of VBC~\cite{rk88}    as well   as RVB
liquids~\cite{ms01,msp02} and a  natural question  is whether QDM  can
describe other phases,  and in  particular  whether they can  describe
phases with {\em gapless   singlet  excitations}.  Motivated by    the
problem  of the spin-$\frac{1}{2}$ KAFH,   we investigated some QDM on
the kagome lattice.   Because of the corner-sharing triangle geometry,
dimer coverings can be handled in a much simpler way (with pseudo-spin
variables~\cite{ez93,msp02}) than on  other lattices.  Exploiting this
property we  introduce  a  QDM  (called $\mu$-model  thereafter)  with
several interesting properties: i) The Hamiltonian allows dimers (from
3 to  6 at a time)  to resonate around  hexagons with amplitudes which
have non-trivial  {\em signs}.  These  signs  are those  arising  when
performing the lowest-order expansion (in  the dimer overlap parameter
introduced      by    Rokhsar       and  Kivelson~\cite{rk88},     see
Sec.~\ref{sec:overlap})  of the KAFH   Hamiltonian in the valence-bond
subspace~\cite{ze95}. These signs are  the crucial difference with the
solvable QDM we introduced  previously~\cite{msp02}.  For  this reason
also quantum  Monte-Carlo simulations  would face the  well-known sign
problem.  ii) In addition to  the topological degeneracy, a feature of
dimer liquids, the ground-state  has a degeneracy which is exponential
with the number of sites,  that  is a {\em extensive  zero-temperature
entropy}.    iii)  The  ground-states   have short-ranged  dimer-dimer
correlations, they are {\em   dimer  liquids}.  We studied the   model
through simple mean-field approximations as well as numerically and we
propose a picture  in which the system  is critical (or at least close
to a critical point).

Because some  parts of the paper  are  relatively independent, we will
now summarize  it so  that  readers may   directly go  to  a  specific
part.  In section~\ref{sec:heisenberg} we  review some  results on the
KAFH model.  Although this  paper is mostly  devoted  to a {\em dimer}
model (sort of extreme quantum limit of the SU$(2)$ spin-$\frac{1}{2}$
model), we find it useful to review  well established facts concerning
the {\em spin} (Heisenberg) model and we motivate  the QDM approach to
the KAFH.  In particular in  Sec.~\ref{ssec:rvb}, we present numerical
results (spectrum and specific heat) obtained by diagonalizing exactly
(on  finite-size   systems)  the   Heisenberg model     restricted  to
nearest-neighbor valence-bonds  subspace.  In  Sec.~\ref{sec:arrow} we
discuss  general properties of  dimer  coverings  on lattices made  of
corner-sharing triangles.   These properties (existence of pseudospins
variables and their dual representation  in terms of arrows) turn  out
to be useful to define  and analyze QDM  on these lattices,  including
kagome.   In  Sec.~\ref{sec:overlap} we explain  the  Rokhsar-Kivelson
overlap expansion when  applied  to the  family  of lattices mentioned
above.  At lowest order the  kinetic energy terms have signs depending
on  the parity   of the  number  of  dimers  involved.   Ignoring  the
amplitudes and keeping only   these sign (Sec.~\ref{sec:qdm}),  we get
kinetic ({\it ie} non-diagonal)  dimer  operators $\mu(h)$ defined  on
every hexagon $h$ of kagome and which realize  an original algebra: i)
$\mu(h)^2=1$, ii) they  anticommute on  neighboring hexagon and   iii)
commute otherwise.  The  rest of the paper is  devoted to the analysis
of  the Hamiltonian  defined as  the  sum  of  all  the $\mu(h)$.   In
Sec.~\ref{sec:1d}  we  start by solving  exactly the  dimer model on a
one-dimensional     lattice.     It  sustains   critical   (algebraic)
correlations and has an  extensive zero-temperature entropy.  Although
we  did not succeed to  find an exact solution to  it, we were able to
show (Sec.~\ref{sec:mukag}) that the kagome $\mu$ model also have such
a zero temperature entropy.   Some mean-field treatments are discussed
in Sec.~\ref{sssec:abc}   and a    competing crystal-like  phase    is
identified.    In  Sec.~\ref{ssec:fermion}   we introduce    fermionic
variables dual  to the $\mu$ operators,  in which the residual entropy
is quite transparent.  This formulation is  reminiscent to that of the
$\mathbb{Z}_2$  gauge   theory in  Ref.~\onlinecite{msp02}.   The last
section  (\ref{sec:numerics}) is devoted  to numerical calculations on
the kagome $\mu$-model.

\section{Some results on the kagome Heisenberg antiferromagnet}
\label{sec:heisenberg}

In this section we review a few results concerning the Heisenberg model
on the kagome lattice. 

\subsection{Classical degeneracy}

The classical kagome antiferromagnet attracted interest because of its
unusual low-temperature properties.   These properties are  related to
the existence of  a local and  continuous degeneracy. Indeed, any spin
configuration    which    as     a vanishing    total    magnetization
$\vec{S}_1+\vec{S}_2+\vec{S}_3=\vec{0}$ on  every  triangle  minimizes
the Heisenberg energy.  Counting {\em planar ground-states} amounts to
find the number of ways one can put $A$, $B$ and $C$ on the lattice so
that each triangle  has spins along  the three different orientations.
This already represents an extensive entropy.\cite{baxter70,hr92} In a
given planar   ground-state one  can look for   closed  loops of  type
$A-B-A-B-\cdots$.   Because  on kagome such a   loop has only $C$-type
neighbors, rotating the spins of  this loop around  the $C$ axis costs
no energy and gives  new (non-planar) ground-states.  Chalker~{\it  et
al.}~\cite{chs92} showed  that all ground-states   can be obtained  by
repeated introduction of such  distortions  into the different  parent
planar states.\footnote{This is in fact a general property of lattices
made of corner-sharing triangles, which  are precisely the lattices we
consider in this paper.} At low temperature this classical spin system
has no magnetic  long-range  order (LRO)  but exhibits  diverging {\em
nematic} correlations when  the   temperature goes to  zero:  although
spin-spin  correlations are short-ranged   the  planes defined by  the
three  spins of  a  triangle are correlated  at  long distances.  This
phenomenon  is a   manifestation of  ``order  by disorder'':   thermal
fluctuations  selects ground-states  with  the largest  number of soft
modes and these are the planar ground-states.

\subsection{Absence of N\'eel long-range order}

It  has been  known for  some  time that the spin-$\frac{1}{2}$ kagome
Heisenberg antiferromagnet has   no N\'eel long-range  order  (LRO) at
zero    temperature.   Early  spin-wave  calculations    by  Zeng  and
Elser~\cite{ze90} indicated that magnetic  order disappears  when going
from   the triangular antiferromagnet  to  the kagome model.  This was
supported by   numerical  calculations of  spin-spin   correlations in
finite  kagome clusters up  to 21 sites~\cite{ze90}.   Two years later
Singh and Huse~\cite{sh92} performed a series expansion about an Ising
limit and came  to the same conclusion about  the absence of  magnetic
LRO.

Although  the classical model has no  N\'eel LRO at $T=0$, the absence
of such order in the spin-$\frac{1}{2}$ case is not completely trivial
because   quantum fluctuations could   select  a  particular type   of
ground-state.  Sachdev~\cite{sachdev92} showed   in the  context of  a
large $N$ expansion  that for a large enough  value of the ``spin'', a
N\'eel LRO sets in (the so called $\sqrt{3}\times\sqrt{3}$ structure).

\begin{figure}
	\begin{center}
	\includegraphics[width=4.5cm,height=4.5cm]{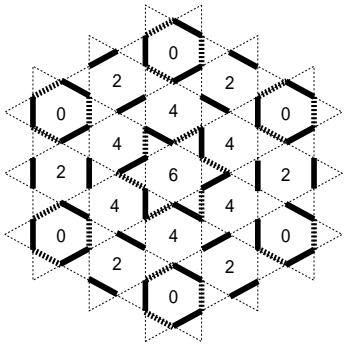}
	\end{center} \caption[99]{  Crystal    of  resonating hexagons
	(marked with  0)  on the kagome  lattice.   Labels $n=0,2,4,6$
	correspond to the  possible  resonance loop $L_n$ around  each
	hexagon                     according                       to
	Table~\ref{tab:loops}. }\label{HexCrystal}
\end{figure}

In 1993 Leung and Elser~\cite{le93} pushed exact diagonalizations to a
36  spins and confirmed the absence  of N\'eel LRO.  They also studied
four-spin correlations (dimer-dimer)  to  investigate the issue  of  a
possible valence-bond crystal (or spin-Peierls, or bond-ordered) phase
made of  resonating hexagons (see  Fig.~\ref{HexCrystal}).  They found
very weak correlations  and suggested the  existence of a liquid phase
(they could  not, however,  definitely  rule out  the possibility of a
very weak  crystalline LRO order).  Nakamura and Miyashita~\cite{nm95}
did Monte-Carlo simulations  including $N=36$ and $N=72$ spins  showed
no kind  of     spin nor   dimer   ordering   down  to $T\simeq    0.2
J$~\footnote{We    use the  following     units  for  the  Hamiltonian
:$\mathcal{H}=2J\sum_{<i,j>}\vec{S}_i\cdot\vec{S}_j$.}    but found  a
low-temperature peak in the specific heat.

On the    analytical side,  Sachdev~\cite{sachdev92}   generalized the
$SU(2)$ model  to an  Sp$(2N)$ symmetry  and  worked  out a  large-$N$
approach based on  bosonic representations.  He  found quantum ordered
phase  with no   broken  symmetries  and unconfined  bosonic  spinons.
However  this result does  not directly explain\footnote{Although the
gauge  degrees  of  freedom  appearing  in  the   Sp$(2N)$ formulation
correspond to singlet excitations, such states are expected to acquire
a (possibly small) gap.} the huge density of low-energy singlet states
that was observed numerically and that we discuss below.

\subsection{Low-energy singlet states}

Lecheminant~{\it   et   al.}~\cite{lblps97}  and   Waldtmann~{\it   et
al.}~\cite{web98} calculated a  large number of low-energy eigenstates
for finite kagome clusters up to 36 sites.  These results pointed to a
large ``residual'' entropy at  low temperatures.  From their data  the
residual entropy per site can be  estimated to be $s_0\sim0.2\log(2)$.
This number was obtained by  counting the number  of eigenstates in  a
finite (and non-extensive) energy window above the ground-state.  This
number was found to scale  as $\sim\alpha^N$ with $\alpha\simeq 1.15$.
The width  of  this  energy  window is  expected  to modify  numerical
prefactors\footnote{First   studies  used    the    value   of     the
singlet-triplet gap (spin gap) as the energy window.   The fate of the
spin gap  in the kagome  Heisenberg  antiferromagnet is not completely
settled, although numerical results suggest  a small but non-zero spin
gap  in the thermodynamic limit.\cite{web98} We  note that a vanishing
spin gap   would cast some  doubt   on  validity the  short-range  RVB
approach   to that spin  model.  Provided   that the  gap  as a smooth
behavior  with the system size,  this does not  prevent one from using
the  gap  as an   energy window  to estimate  the   entropy.}  but not
$\alpha$  which  is  directly   related to   the an   entropy per site
$s_0=\log{(\alpha)}$.

\subsection{Specific heat}

The  entropy change   between  zero and   infinite temperature  can be
extracted from the specific heat $c_v(T)$.  The first high-temperature
(HT) series expansion for the  kagome antiferromagnet was carried  out
by Elstner and Young~\cite{ey94}.  This approach showed a huge entropy
deficit             of               about         40\%              :
$\int_0^\infty{c_v(T)/TdT}\simeq0.6\ln(2)$.   However,   this   direct
evaluation of the specific heat from HT series is  not accurate at low
temperatures  and they concluded  the possibility of a low-temperature
peak  in  the  specific heat.   A  quantum   Monte-Carlo simulation by
Nakamura and Miyashita~\cite{nm95} also  found a low-temperature peak.
Such a peak was also found by a decimation calculation~\cite{tr96}.  A
recent    exact    diagonalization    work    by  Sindzingre~{\it   et
al.}\cite{smlb00} also found such a peak in a 36-sites sample.
An improved method  of calculation  of $c_v(T)$ from  high-temperature
series  expansion, which   is  quantitatively  accurate  down to   zero
temperature in  most frustrated  magnets~\cite{bm00} shows that  about
20\%   of  the  total    entropy   is still     missing  at very   low
temperatures~\cite{bm01},  in   agreement with  exact diagonalizations
data.

\subsection{About residual entropies}

If a  system has a  number of  states growing  exponentially (with the
system size) in a non-extensive energy  window above the ground-state,
it as a  extensive residual entropy  at zero  temperature.   In such a
case, although the ground-state can be unique on finite systems, it is
in  fact   exponentially     degenerate    in   the    thermodynamic
sense\footnote{Take  the   infinite volume   limit  before taking the
temperature  to zero.}.  One can  construct some simple models with an
extensive residual   entropy   (the  Ising  antiferromagnet  on    the
triangular lattice for instance)  but it is  usually lifted  by almost
any infinitesimally small perturbation.  An extensive entropy at $T=0$
is not a generic situation  but instead requires  some fine tuning (to
zero) of all  these perturbations.  For these  reasons we think it  is
unlikely that the spin-$\frac{1}{2}$ kagome Heisenberg antiferromagnet
has    a  $T=0$   residual   entropy.     Consider   some  Hamiltonian
$\mathcal{H}(\lambda)=\mathcal{H}_0+\lambda\mathcal{H}_1$        where
$\mathcal{H}_0$ has an  exponential  ground-state degeneracy which  is
lifted by $\mathcal{H}_1$.  At small  $\lambda$ the specific heat  may
have     low temperature peak     which  entropy  corresponds  to  the
ground-state   degeneracy  of  ${\mathcal  H}_0$.     Upon taking   the
$\lambda\to0$ limit,  the temperature  of  the  peak goes  to  zero as
well. This  is the picture we  have in mind for the spin-$\frac{1}{2}$
kagome antiferromagnet  and this paper  discuss a possible scenario in
which the  role of $\mathcal{H}_0$ is played  by a quantum dimer model
(defined as the $\mu$-model in section~\ref{sec:qdm}).

\subsection{Resonating valence-bond subspace}
\label{ssec:rvb}

Quantum  dimer models can   provide   effective descriptions of   some
magnetically disordered phases  of antiferromagnets.  We first wish to
motivate   the  restriction  of  the   spin  Hilbert   space   to  the
first-neighbor resonating valence-bond (RVB)  subspace which has  been
used in a number  of  works~\cite{ez93,ze95,m98,mm00} for the   kagome
problem.   This space is generated   by all valence-bond states  where
spins   are   paired    into first-neighbor    singlets   (dimers   or
valence-bond).  Because spin-spin  correlations are very short-ranged,
it  is rather natural to consider  the ground-state wave function as a
linear superposition of valence-bond  states.  The crucial point is to
understand whether   valence bonds  beyond first  neighbors  should be
included or not in the Hilbert space  to (qualitatively) get the right
physics.   We  will only  partially address it   in  this paper.  This
first-neighbor RVB  limit is   the  simplest  subspace  which  has  an
exponential number  of states that  could explain the proliferation of
low-energy singlets observed  numerically.  In addition, this subspace
provides a     reasonably  good   variational  energies.   Zeng    and
Elser~\cite{ze95}   and Mambrini  and   Mila~\cite{mm00} computed  the
ground-state energy of  the  Heisenberg Hamiltonian restricted to  the
first-neighbor RVB subspace\footnote{Zeng and  Elser did  not consider
all dimer  coverings  but restricted their   study to one  topological
sector.  This has  no consequence  in the  thermodynamic but it  is an
approximation on  finite size systems.   For this reason their results
differ from those  of Mambrini and Mila who  worked in the full  dimer
space.}.  For   a      sample   of    36  sites    their        result
($2<S_i{\cdot}S_j>=-0.4218$)  is   $3.8$\%    higher  than  the  exact
ground-state energy  obtained~\cite{le93,web98}    in  the  full  spin
Hilbert space ($2<S_i{\cdot}S_j>=-0.4384$).   Zeng and Elser were able
to improve  significantly  this  variational   estimate by a    simple
optimization of the dimer wave function in the vicinity of each defect
triangle, but without changing the  dimension of the Hilbert space. To
our    knowledge this ``optimized dimerizations   basis''  is the best
variational one for the kagome problem.   It is also worth saying that
in the fermionic  large-$N$  extension of the Heisenberg  model, first
neighbor valence-bond states  arise as degenerate ground-states in the
$N\to\infty$ limit.\cite{r90} $1/N$ corrections  will then introduce a
dynamics among  these dimerized  states.  Marston and Zeng~\cite{mz91}
used such a fermionic $SU(N)$ extension of the Heisenberg model on the
kagome  lattice and found that such  $1/N$ corrections could favor the
crystal       of      resonating     hexagons     mentioned      above
(Fig.~\ref{HexCrystal}).

\begin{figure}
	\begin{center}
	\includegraphics[width=7cm,bb=18 430 592 718]{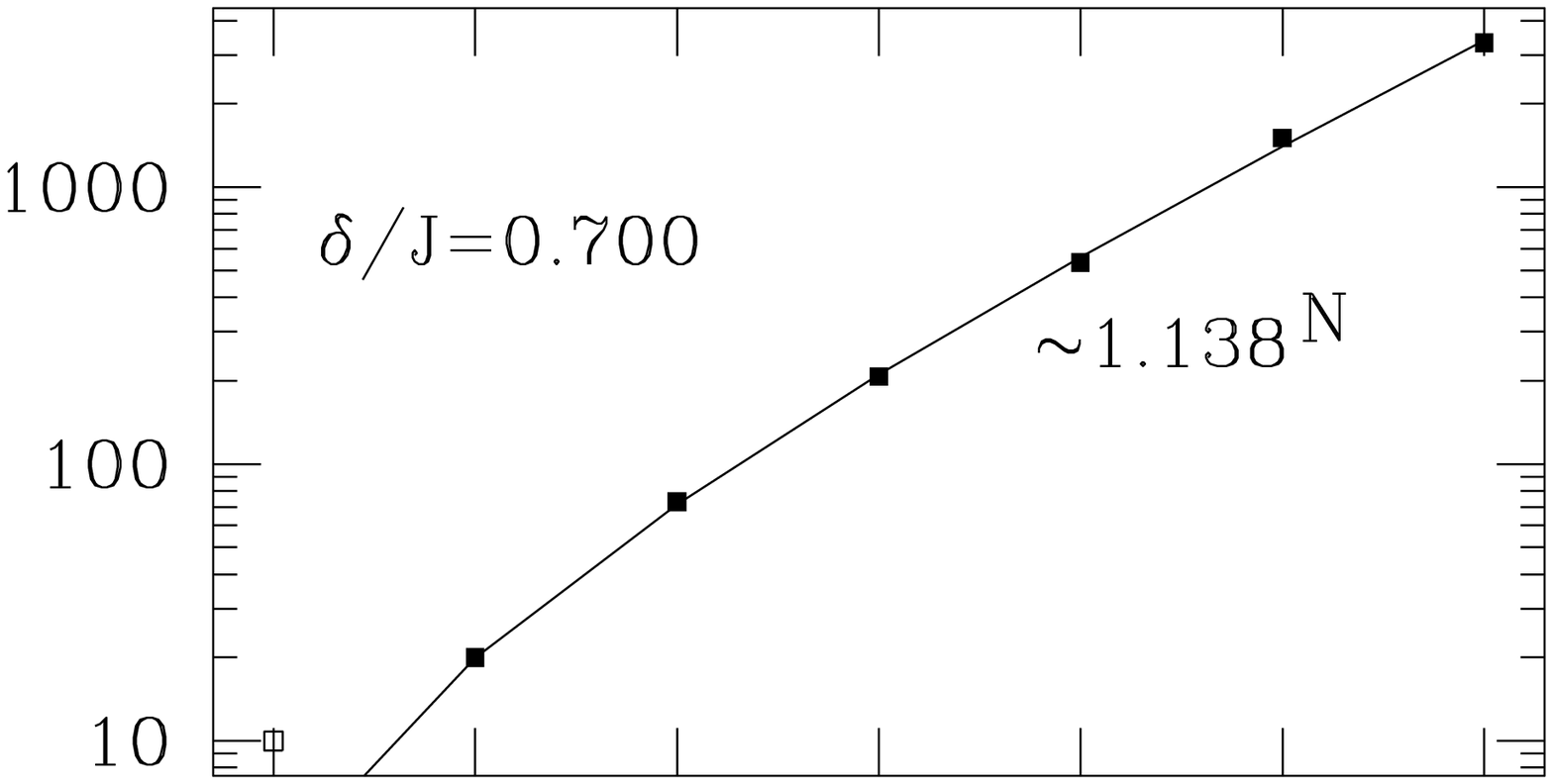}\\
	\includegraphics[width=7cm,bb=18 345 592 718]{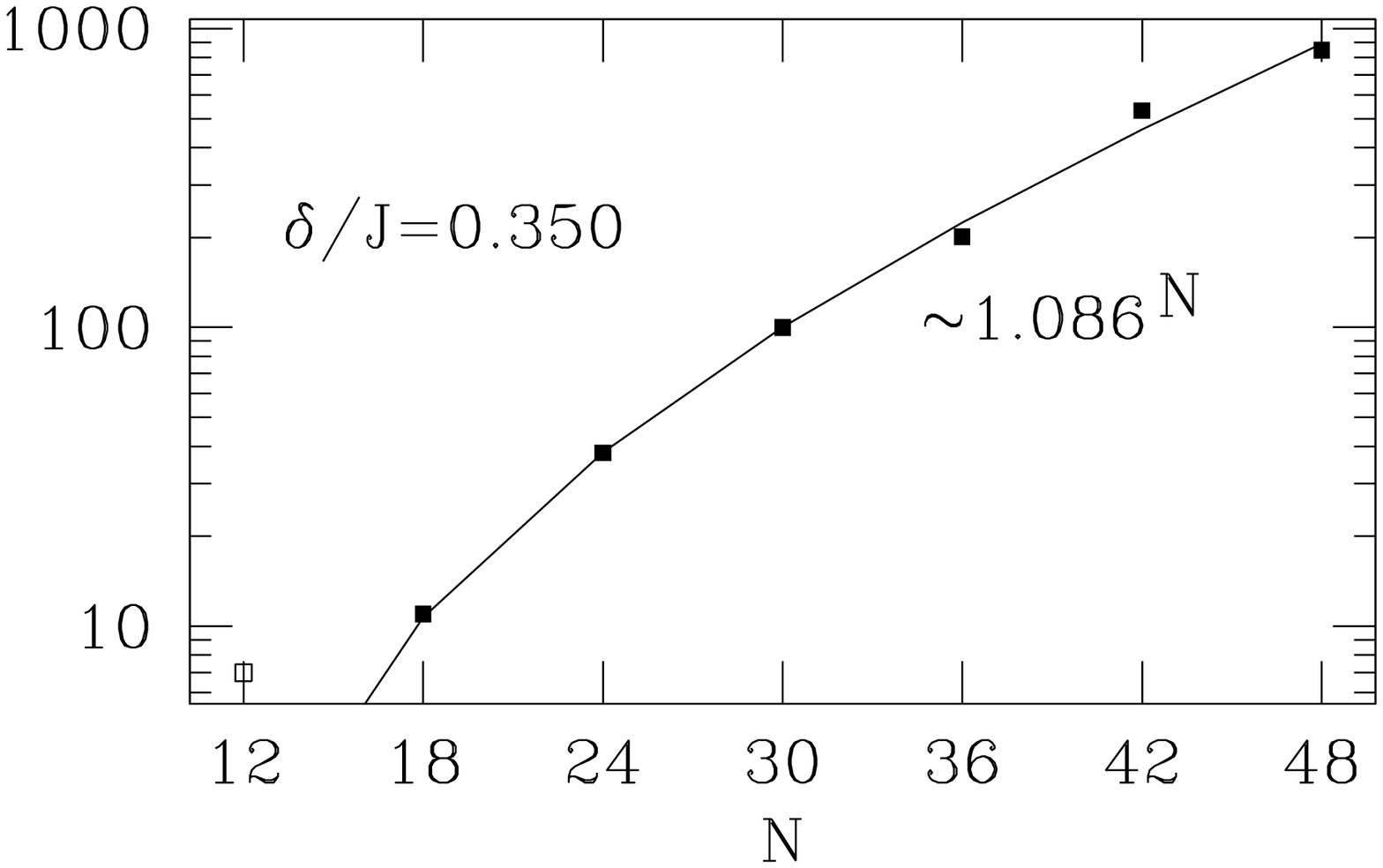}
	\end{center}

	\caption[99]{Kagome Heisenberg antiferromagnet diagonalized in
	the  first-neighbor   RVB  space.   An exponential   number of
	eigenstate    is      observed  in     the    energy    window
	$[E_0,E_0+\delta]$,  where    $E_0$   is   the    ground-state
	energy. The results for two values of  $\delta$ are shown. The
	full  lines are  quadratic  least-square fits to  the data for
	$N\ge18$.    The      definition  of  $J$      is   such  that
	$\mathcal{H}=\sum_{<i,j>}\vec{S}_i\cdot\vec{S}_j$. }

	\label{states_below_delta}
\end{figure}

\begin{figure}
	\begin{center}
	\resizebox{7cm}{!}{\includegraphics{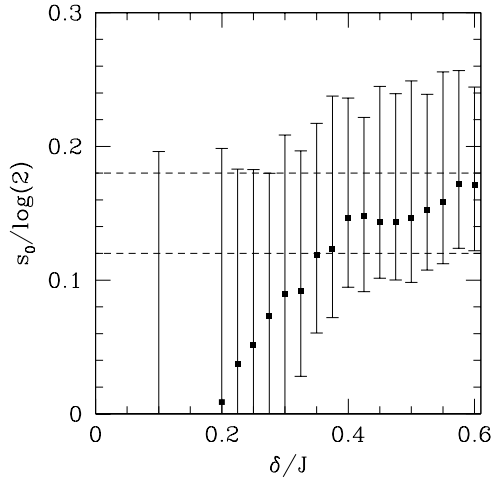}}
	\end{center}  \caption[99]{For   each  width  $\delta$ of  the
	energy window we fit the exponential increase of the number of
	energy levels to  estimate the  zero-temperature entropy $s_0$
	(as  in  Fig.~\ref{states_below_delta}).  Error bars come from
	the uncertainty of the least-square fits.}\label{s0_of_delta}
\end{figure}

A last argument   for the first-neighbor  RVB approach  to the  kagome
problem is the fact that the spectra of the Heisenberg model projected
into this subspace  reproduce a continuum of singlet  states as in the
case of  spectra computed in the   full spin Hilbert  space.  This was
first noticed  by Mambrini and  Mila~\cite{mm00} on  samples up to  36
spins    and  we   extended  their   study    to samples     up to  48
sites. Figure~\ref{states_below_delta} shows the exponential number of
low-energy states in a  finite energy window  $[E_0,E_0+\delta]$ above
the ground-state.  We   analyzed  this  exponential  proliferation  of
energy levels as a function of the system size {\em and} as a function
of  the  energy window.  Although we  have  7 complete  spectra  up to
$N=48$ sites the dependence on the width of the energy window makes it
difficult  to   give an  precise   estimate  of  the low   temperature
entropy.  For  each value  of $\delta$  we  plot the logarithm  of the
number of states in the window as in Fig.~\ref{states_below_delta}.  A
(least-square) fit  is performed  to extract  the  leading exponential
behavior  when $N\to\infty$.   Error  bars are obtained  in a standard
way.\footnote{We repeat the fit when one and  when any two data points
are removed.  The minimum  and  maximum values  of the coefficient are
used to    estimate the uncertainty.}    In  principle  this procedure
measures   the zero-temperature  entropy   provided  that $\delta/J\ll
N$.  The result is  summarized Fig.~\ref{s0_of_delta}.  Unfortunately,
one cannot use  too small  values  of $\delta$ because  discretization
effects  scatter the data when $\delta$  is of  the  same order as the
typical level spacing in the smallest samples.  This is the reason for
the  increasingly large error  bars we  obtain  when $\delta$ is below
$0.6\sim0.8J$  (Fig.~\ref{s0_of_delta}).  However, from these  results
it appears likely that   a significant part   of the total entropy  is
present at temperature much lower than the  energy scale $J$.  Indeed,
the  values  of  $s_0$  compatible  with  the  set  of data  displayed
Fig.~\ref{s0_of_delta}  is   $0.1\log(2)\le{s_0}\le0.2\log(2)$.   Only
these  values are within    the  error bars   of  all estimates   from
$\delta=0.4J$ to $\delta=1.2J$.

\begin{figure}
	\begin{center}
	\resizebox{7cm}{!}{\includegraphics{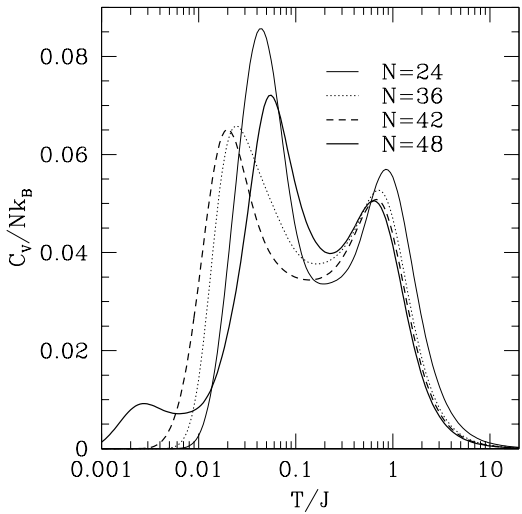}}
	\end{center} \caption[99]{Specific heat per site of the kagome
	Heisenberg model    restricted   to   the  first-neighbor  RVB
	subspace.  The lowest  peak  at  $T/J\simeq3.10^{-3}$  ($N=48$) is
	a finite-size artifact.  }  \label{CvKagRVB}
\end{figure}

Computing the specific heat is  another method to  look for a possible
residual  entropy.   In    the case of     the  kagome antiferromagnet
diagonalized in the full spin Hilbert space a low-temperature peak was
observed~\cite{smlb00},  as well as  in  some experiment on a spin-3/2
kagome compound~\cite{ramirez00}.   From its   low sensitivity  to  an
applied   magnetic field,  this     peak was attributed   (mostly)  to
non-magnetic  singlet states.  In this work   we computed the specific
heat of the kagome antiferromagnet in the first-neighbor RVB subspace.
This calculation is done    from  the spectra obtained   by  numerical
diagonalizations up  to $N=48$ sites.  The  results shown are shown in
Fig.~\ref{CvKagRVB}.  The   maximum of   $C_V(T)$ around  $T=0.7J$  is
almost converged  to its thermodynamic  limit.  It corresponds  to the
onset  of short-range   correlations.   For all  sample sizes  a large
low-temperature peak is present at   or below $T=0.07J$.  It is  still
size-dependent  but its  entropy  roughly corresponds  one half of the
total entropy of  the model (the  total entropy  per  site of the  RVB
space    is  $\frac{1}{3}log(2)$),   in   agreement  with results   of
Fig.~\ref{CvKagRVB}.  The similarity  between these  results and those
obtained in the full spin Hilbert space is another  support to the RVB
approach.

To summarize, we have reviewed  several arguments indicating that  the
unusual  low-temperature peak  in the   specific  heat of  the  kagome
antiferromagnet  might   be explained within  the   framework of a RVB
space.  We would like to conclude this  section by mentioning that the
spin-$\frac{1}{2}$   Heisenberg model  may  have  a  large  number  of
low-energy  singlet states  on  other lattices made  of corner-sharing
triangles.   This was  observed  numerically~\cite{waldtmann00} on the
frustrated three-leg ladder shown Fig.~\ref{chain}.  We will come back
to that  model  in  Sec.~\ref{sec:1d}.  In  Ref.~\onlinecite{sg01} the
squagome lattice was introduced and some low-energy states reminiscent
of the   kagome  ones were identified   in a  large-$N$   approach.  A
decimation  method  applied   to  this  lattice and    also predicts a
low-temperature peak in the specific heat of the model~\cite{tomczak}.
A numerical diagonalization study of the Heisenberg antiferromagnet on
the Sierpinski gasket~\cite{sierp} found a low-temperature peak in the
specific heat as well.

\section{Dimer coverings on lattices made of corner-sharing triangles}
\label{sec:arrow}

Before studying  the restriction of the Heisenberg  spin model  to the
valence-bond subspace  we   will introduce   some properties of   dimer
coverings   on lattices  made of  corner-sharing  triangles (including
kagome).    A very   useful    property   discovered  by Elser    and
Zeng~\cite{ez93} is that dimer coverings on these  lattices can be put
in    one-to-one   correspondence   with   configurations   of   arrow
variables. Also,  this representation is  intimately connected  to the
existence of (Ising like) pseudo-spin variables.\cite{msp02}

\begin{figure}
	\begin{center}
	\resizebox{5cm}{5cm}{\includegraphics{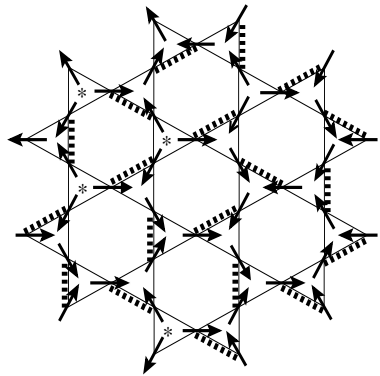}}
	\end{center} \caption[99]{Arrow  representation  of   a  dimer
	coverings.}\label{arrow}
\end{figure}

The correspondence  between dimer coverings  on the kagome lattice and
{\em sets of  arrows} is illustrated  in  Fig.~\ref{arrow}. Each arrow
has two possible directions: it must  point toward the interior of one
of the  two  neighboring triangles.  In a triangle, a dimer connect
two sites where the arrows point inwards. 
In a defect triangle (without any dimer, marked with $*$ in
Fig.~\ref{arrow}),   the three arrows point outwards.  Therefore, at each
triangle there is a constraint imposing that the number of incoming
arrows is even.

Dimer moves translate very  simply  in the arrow representation.   One
can easily verify  that $\sigma^x(h)$ (see Ref.~\onlinecite{msp02} and
appendix~\ref{sec:ZE})  does   nothing  but  flipping  the $6$  arrows
sitting around hexagon $h$ and  that such  an operation conserves  the
constraint  for all  triangles.    Any dimer move\footnote{This  move
should  not, however, contain topologically  non-trivial loops.}  is a
product $\sigma^x(h_1)\sigma^x(h_2)\cdots$ where  $h_1,h_2,\cdots$ are
the hexagons enclosed in the loops.   This operation successively flips
all the arrows around $h_1,h_2,\cdots$.  The result does not depend on
the order in  which hexagons  are flipped  so  the $\sigma^x$ operators
obviously commute in this language.

\subsection{Medial lattice construction}
\label{ssec:medial}
\begin{figure}
	\begin{center}
	\resizebox{4cm}{4cm}{\includegraphics{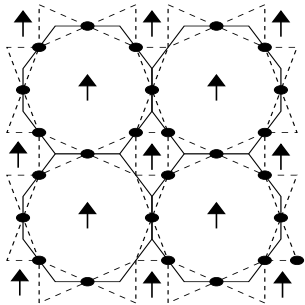}}
	\end{center}
	 \caption[99]{Squagome lattice (dashed lines)
	as     the medial lattice    of   the octagonal lattice.}
	\label{Oct2Squag}
\end{figure}

The arrow   representation (as   well   as the  pseudospin   operators
$\sigma^x$   and      $\sigma^z$     introduced     by    Zeng     and
Elser~\cite{ez93,ze95}          -- see  appendix~\ref{sec:ZE}) can  be
generalized  to all lattices made by corner-sharing triangles.  
The  kagome case is the simplest  example in two dimension, 
another being the squagome
lattice~\cite{sg01} (Fig.~\ref{Oct2Squag}). The Sierpinski 
gasket ~\cite{sierp} is an example of fractal structure of dimension
between one and two also made by corner-sharing triangles.

\begin{figure}
	\begin{center}
	\resizebox{4cm}{4cm}{\includegraphics{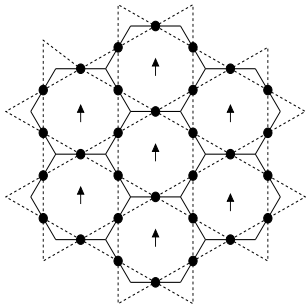}}
	\end{center}  \caption[99]{Kagome  lattice (dashed lines   and
	black dots) constructed as the medial lattice of the hexagonal
	(full lines)   lattice.  The location of  the  pseudospins are
	indicated by up spins.}
	\label{Hex2Kag}
\end{figure}

\begin{figure}
	\begin{center}
	\resizebox{!}{1.5cm}{\includegraphics{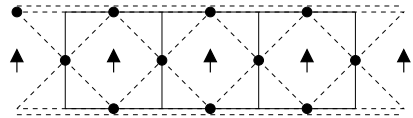}}
	\end{center}

	 \caption[99]{A  frustrated    three-spin      ladder  (dashed
	 lines)obtained as the medial  lattice of  (trivalent) two-leg
	 ladders.}  \label{chain}
\end{figure}

\begin{figure}
	\begin{center}
	\includegraphics[width=7cm,height=1.7cm]{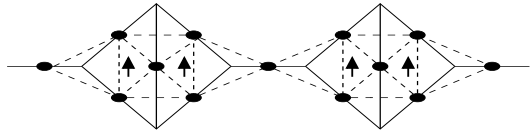}
	\end{center}
	 \caption[99]{Another example of  chain.}
	\label{chain2}
\end{figure}

\begin{itemize}
	\item[a)] Start with a    trivalent  lattice $H$, that is    a
	lattice  where each site  has three  neighbors  (full lines in
	Figs.~\ref{Hex2Kag},     \ref{Oct2Squag},     \ref{chain}  and
	\ref{chain2}.  The hexagonal lattice (Fig.~\ref{Hex2Kag})   is
	the  simplest two-dimensional  example.

	\item[b)] Construct its {\em medial lattice} $K$: Sites of $K$
	are,  by  definition, the centers  of the  links  of $H$.  The
	sites sitting on the three links of  $H$ connected to the same
	site of $H$ are connected together.  The medial lattice of the
	hexagonal   lattice is the     kagome lattice.  Since $H$   is
	trivalent, $K$ is made  of  corner-sharing triangles.

	\item[c)] Associate a pseudospin to each  plaquette of $H$ (ie
	to hexagons of the kagome lattice in the example).
\end{itemize}

In the  following, we will  use $N$ for  the  number of sites  in $K$,
which is equal to the number of links in $H$.  The  number of sites in
$H$ will be $2N/3$, which is equal to the number  of triangles in $K$.
The  number of plaquette  (or faces) in $H$  is equal to the number of
pseudospins,  we write it  $N_{\rm{ps}}$. For two-dimensional cases we
can apply     Euler's relation   to  the  lattice    $H$  and  we find
$(2N/3)-N+N_{\rm{ps}}=2-2g$ where $g$ is its genus ($g=1$ for a torus,
$g=0$ for a sphere).

\begin{figure}
	\begin{center}
	\resizebox{3cm}{!}{\includegraphics{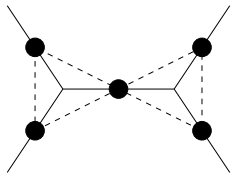}}  \end{center}
	\caption[99]{Medial  lattice  construction. Starting  from   a
	trivalent  lattice (full lines)   we construct a lattice which
	sites (black dots)  are  centers of  the links.   The sites of
	this  new lattice are  linked together  (dashed lines) to form
	triangular plaquettes.}\label{medial}
\end{figure}

\subsection{Counting dimer coverings with arrows}

The number of  dimer coverings of  any lattice of type $K$ (including,
for  instance,  the   one-dimensional  examples  of  Figs.~\ref{chain}
and~\ref{chain2}) is
\begin{equation}
	\mathcal{N}_{\rm dim. coverings}=2^{N/3+1}
	\label{eq:Ndcov3}
\end{equation}
This result can be obtained with  the arrow representation. Each arrow
has two  possible directions, which  give $2^N$ states.  The fact that
there can  only  be  $0$ or  $2$ incoming   arrows  for  each triangle
introduces one constraint per triangle. There are $2N/3$ triangles but
only  $2N/3-1$   constraints are  really   independent,  which   gives
Eq.~\ref{eq:Ndcov3}.

The fact that $2N/3-1$ constraints are independent can be checked with
the following argument.  We focus on the  trivalent lattice $H$ on the
bond of which the  arrows live.  First  transform $H$  into a  tree by
recursively  cutting every bond  that does not  disconnect the lattice
into two parts.   The final tree is  still trivalent so the number  of
leafs $L$ is related to the  number of vertices $V$  by $L=V+2$.  Each
bond gives  two leafs when  it is cut so that  $L/2$  is the number of
cuts. One can now set  the arrows directions  on the leafs.  There are
$L/2$  such independent arrows.  Using   the constraints associated  to
each vertex, the arrows are then determined  on all the other bonds of
the tree by progressively going from the leafs  toward the root. It is
simple to check that the  last constraint encountered when reaching the
root   is   automatically    satisfied.     From    this  we    obtain
$2^{L/2}=2^{V/2+1}$ dimer   configurations,   which is  equivalent  to
Eq.~\ref{eq:Ndcov3} since $3V=2N$.

\section{From spins to dimers: overlap expansion}
\label{sec:overlap}
When restricted to the RVB  subspace, the  Heisenberg model induces  a
complicated  dynamics  on valence-bonds. This   dynamics is intimately
related to the non-orthogonality  of these valence-bond states that we
describe below.

\subsection{Scalar product and loops}

\begin{figure}
	\begin{center}
	\includegraphics[width=5cm]{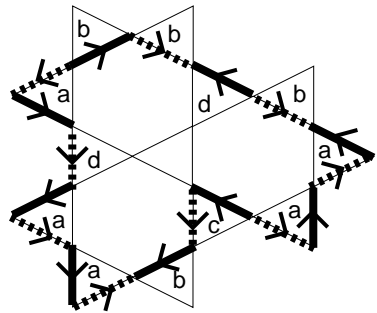}
	\end{center}
	\caption[99]{Transition graph between two dimer coverings. The
	bonds are   oriented so that all  hexagons  are clockwise. The
	loop passes through four types of  triangles: a,b,c and d (see
	text).}\label{fig:scalsign}

\end{figure}

The scalar product  of two  valence-bond states  can be  computed from
their transition  graph\cite{s88}  (loop covering obtained  by drawing
both dimerizations on  the top of  each other).  We  first need a sign
convention for valence-bond states.  A  simple choice is to orient all
the bonds so that all hexagons  are clockwise\footnote{A dimer on bond
$ij$   oriented     from   $i$   to      $j$  then    corresponds   to
$(\left|\uparrow_i\downarrow_j\right>-\left|\uparrow_j\downarrow_i\right>)/\sqrt{2}$.} 
(see Fig.~\ref{fig:scalsign}). With this choice, the scalar product of
two valence-bond states $\left|a\right>$ and $\left|b\right>$ is
\begin{equation}
	\left<a | b\right> =
		\prod_{Loops} \left[
		\left(1/2\right)^{L/2-1}
		\left(-1\right)^{1+N_{\rm hex}+L/2} 
		\right]
		\label{eq:scal}
\end{equation}
where the product runs over non-trivial (of  length $>2$) loops in the
transition  graph of $a$  and $b$, $L$ is  the length  of the loop and
$N_{\rm hex}$   is the number  of hexagons  enclosed  by a  loop.  For
instance, the    loop displayed  Fig.~\ref{fig:scalsign}  has  $N_{\rm
hex}=3$.  The   factor $\left(1/2\right)^{L/2-1}$ in Eq.~\ref{eq:scal}
is   valid  on        any      lattice   whereas      the        signs
$\left(-1\right)^{1+N_{\rm{hex}}+L/2}$  are     associated    to   the
corner-sharing  triangle  geometry\footnote{Notice the difference with
bi-partite lattices where  it is possible to  orient the bonds so that
the sign $\left<a|b\right>$  is always be  positive~\cite{s88}.}.  The
sign of $\left<a|b\right>$ would   just be $\left(-1\right)^{L/2}$  if
all  the  bonds  dimers  were oriented   clockwise   around the loops.
Consider  the  triangles on which a   loop  passes.  We classify these
triangles in four types ($a$, $b$, $c$  and $d$) as follows. Some have
{\em   two} edges    on    the   loop     (types $a$   and   $c$    in
Fig.~\ref{fig:scalsign}) and the others have only {\em one} (types $b$
and $d$).  Some are {\em inside} the loop  (types $a$ and $d$) and the
others are  {\em outside} ($b$ and  $c$).  Triangles of  type $a$ give
two anticlockwise bonds and the $(-1)$  factors cancel. Every triangle
of type $b$ (resp.   $c$) give one (resp.   two) clockwise bond(s) and
do  not contribute to  the sign.  Each  triangle  of type $d$ gives an
anticlockwise  bond and contributes by  a  factor $(-1)$ to the scalar
product.  Let $N_d$ be the  number of such  triangles.  So far we have
shown                                                             that
${\rm{sign}}\left[\left<a|b\right>\right]=\prod_{Loops}\left(-1\right)^{N_d+L/2}$. Using
the arrow representation (and the associated  constraint) one can show
that $(-1)^{N_d}=(-1)^{N_{\rm hex}+1}$.  The argument - not reproduced
here for brevity - uses the  fact the parity of  arrows coming out and
in of a given  cluster of sites is  fixed by the  number of links  and
sites in that  cluster, an thus to  the number of hexagons enclosed by
the loop.

Rokhsar and Kivelson~\cite{rk88}  generalized the scalar product given
by Eq.~\ref{eq:scal} by giving two arbitrary fugacities $x$ and $y$ that couple to
the number and to the length of the loops:
\begin{equation}
	\left<a | b\right>_x = \Omega_{x\;a,b} =
		\prod_{Loops} \left[
		y x^{L/2}
		\left(-1\right)^{1+N_{\rm hex}+L/2} 
		\right]
		\label{eq:scalx}
\end{equation}
$x$ and $y$ can be considered as  formal expansion parameters ($x=1/2$
$y=2$ in the  physical spin-$\frac{1}{2}$ case).   The choice $y=2$ is
usually adopted in  the literature~\cite{rk88,ze95} but  we will keep
$y$ explicit so that other cases may  be considered.  When $x\to0$ the
overlap matrix $\Omega_x$ becomes diagonal.

\subsection{Rokhsar  and Kivelson  scheme}

When restricted to the   first-neighbor   RVB space, the    Heisenberg
Hamiltonian induces a  dynamics on valence-bonds.   These valence-bond
states are not orthogonal so we have a generalized eigenvalue problem.
Orthogonalized  valence-bond  states  $\left|\tilde{a}\right>$  can be
obtained~:
\begin{eqnarray}
	\left|\tilde{a}\right>&=&
		\sum_{b} (\Omega_x^{-1/2})_{ab} \left|b\right>
\end{eqnarray}
and the matrix to diagonalize is
\begin{eqnarray}
	\mathcal{H}^{\rm eff}_{ab}(x) &=&
			\left<\tilde a |\mathcal{H}| \tilde{b} \right>_x \\
			&=& \sum_{a'b'}
			(\Omega_x^{-1/2})_{aa'}
			\left<a'| \mathcal{H} |b'\right>_x
			(\Omega_x^{-1/2})_{bb'}
\end{eqnarray}
where        $\mathcal{H}$    is   the   Heisenberg        Hamiltonian
$\mathcal{H}=\sum_{<ij>}\vec{S}_i\cdot\vec{S}_j$.   From  now  we will
only   deal  with      {\em   orthogonalized}   valence-bond    states
$\left|\tilde{a}\right>$ but  we   will drop the tilde    for clarity.
$\mathcal{H}^{\rm eff}(x=1/2)$ was  diagonalized numerically to obtain
the results of Fig.~\ref{states_below_delta} and
\ref{CvKagRVB} (see also Refs.~\onlinecite{ze95} and
\onlinecite{mm00}).    $\mathcal{H}_{\rm{eff}}(x>0)$  is non-local and
many dimers can hop  simultaneously to quite different configurations.
However,  since $x=\frac{1}{2}<1$, the  tunneling probability for such
events decrease exponentially  with  the  loop  length.   Up to  order
$x^n$, $\mathcal{H}^{\rm eff}(x)$ is   local and only   contains terms
with $\le  n$ dimers.  Following  Rokhsar and  Kivelson's work  on the
square lattice~\cite{rk88}, Zeng and Elser~\cite{ez93,ze95} considered
the  small-$x$  expansion  of $\mathcal{H}^{\rm{eff}}$  on the  kagome
lattice up to order $x^{6}$. Up to a constant we have~:
\begin{eqnarray}
	\mathcal{H}^{\rm eff }(x)= - \sum_{h,\alpha}
	h_\alpha(x)
	\left|L_\alpha \right>\left< \bar{L}_\alpha\right|
	+ {\rm h.c.}
	+\mathcal{O}(x^{6})
	\label{eq:heffkag}
\end{eqnarray}
where the sum on $h$ runs over all hexagons, the  sum on $\alpha$ runs
over all the loops enclosing  a that hexagon. The tunneling amplitudes
$h_\alpha(x)$ are given  in Tab.~\ref{tab:loops}.
These result reduces
to Zeng and Elser's~\cite{ze95}  when  $y=1/x=2$.  Notice that   other
terms of order $x^6$ exist and  involve 6-dimer moves around {\em two}
hexagons.

Eq.~\ref{eq:heffkag}  can be   obtained   from   the scalar    product
formula.   The latter   is  valid   for   any lattice   $K$   made  of
corner-sharing   triangles (see     section~\ref{ssec:medial})     and
Eq.~\ref{eq:heffkag}  can   be generalized   to  these lattices.   Now
hexagons are  replaced by plaquettes of  the  lattice $H$.  Consider a
loop $\alpha$ encircling a single plaquette.  It has  a length $L$ and
encloses $N_t$   triangles. The the   amplitude $h_\alpha(x)$ for that
dimer move is
\begin{eqnarray}
	h_\alpha(x) &=& \frac{1}{2}y(-x)^{L/2} \left(L-yN_t\right)
	+\mathcal{O}(x^L_{\rm min})
\end{eqnarray}
where $L_{\rm min}$ ($=6$ for kagome) is the size of the plaquettes of
$H$.   Unlike  the   square~\cite{rk88}   or the triangular    lattice
case~\cite{ms01}, {\em no sign convention  for the dimer coverings can
turn the signs of the amplitudes $h_\alpha$ all equal}.

\begin{table}
\begin{center}
\begin{tabular}{|c|c|c|c|c|c|c|}
\hline
	Loop $\alpha$ & $L/2$ & $h_\alpha(x)$ &~~$\sigma^x$~~ & ~~$\mu$~~ & ~~$\tilde{\mu}$~~ \\
	\hline
	\begin{picture}(50,34)(-24,-15)
	\pA{\La}\pB{\Lb}\pC{\Lc}\pD{\Ld}\pE{\Le}\pF{\Lf}\KagHex
	\end{picture} & 3 & $-3yx^3=-3/4$ & + & - & +\\
	\hline
	\begin{picture}(50,34)(-24,-13)
	\pA{\La}\pB{\La}\pG{\Lc}\pC{\Lc}\pD{\Ld}\pE{\Ld}\pJ{\Lf}\pF{\Lf}
	\KagHex\pG{\C}\pJ{\C}
	\end{picture} & 4 & $yx^4(4-y)=1/4$ & + & + & +\\
	\begin{picture}(50,34)(-24,-13)
	\pA{\La}\pB{\La}\pG{\Lc}\pC{\Lc}\pD{\Ld}\pE{\Le}\pF{\Le}\pK{\La}
	\KagHex\pG{\C}\pK{\C}
	\end{picture} & 4 & $yx^4(4-y)=1/4$ & + & + & -\\
	\begin{picture}(50,34)(-24,-13)
	\pA{\La}\pB{\La}\pG{\Lc}\pC{\Lb}\pH{\Ld}\pD{\Ld}\pE{\Le}\pF{\Lf}
	\KagHex\pG{\C}\pH{\C}
	\end{picture} & 4 & $yx^4(4-y)=1/4$ & + & + & +\\

	\hline
	\begin{picture}(50,34)(-24,-13)
	\pA{\La}\pB{\La}\pG{\Lc}\pC{\Lb}\pH{\Ld}\pD{\Ld}
	\pE{\Ld}\pJ{\Lf}\pF{\Le}\pK{\La}
	\KagHex\pG{\C}\pJ{\C}\pK{\C}\pH{\C}
	\end{picture} & 5 & $yx^5(2y-5)=-1/16$ & + & - & +\\
	\begin{picture}(50,44)(-24,-23)
	\pA{\Lf}\pB{\Lb}\pL{\Lb}\pC{\Lb}\pH{\Ld}\pD{\Ld}
	\pE{\Ld}\pJ{\Lf}\pF{\Le}\pK{\La}
	\KagHex\pL{\C}\pJ{\C}\pK{\C}\pH{\C}
	\end{picture} & 5 & $yx^5(2y-5)=-1/16$ & + & - & -\\
	\begin{picture}(50,44)(-24,-23)
	\pA{\Lf}\pB{\La}\pL{\Lb}\pC{\Lc}\pG{\Lc}\pD{\Ld}
	\pE{\Ld}\pJ{\Lf}\pF{\Le}\pK{\La}
	\KagHex\pL{\C}\pJ{\C}\pK{\C}\pG{\C}
	\end{picture} & 5 & $yx^5(2y-5)=-1/16$ & + & - & +\\
	\hline
	\begin{picture}(50,50)(-24,-23)
	\pA{\Lf}\pB{\La}\pL{\Lb}\pC{\Lb}\pG{\Lc}\pD{\Lc}
	\pE{\Ld}\pJ{\Lf}\pF{\Le}\pK{\La}\pH{\Ld}\pI{\Le}
	\KagStar
	\end{picture} & 6 & $yx^6(6-3y)=0$ & + & + & +\\
	\hline
\end{tabular}
\end{center}

\caption{The 8 different classes of loops which can surround an hexagon.
	Including all   possible  symmetries   we  find   32  possible
	loops. $L/2$ is  the number of  dimers involved and $h_\alpha$
	is  the tunneling amplitude  (at lowest order) associated with
	each loop in a small-$x$ expansion of  the Heisenberg model in
	the RVB space (see text). The value for the physical case $y=1/x=2$
	is given.}

\label{tab:loops}
\end{table}

\section{Quantum dimer Hamiltonian}
\label{sec:qdm}

In Ref.~\onlinecite{msp02} we introduced a solvable model:
\begin{equation}
	\mathcal{H}_0= - \sum_{h,\alpha}
	\left|L_\alpha \right>\left< \bar{L}_\alpha\right|
	+ {\rm h.c.} \\
	= - \sum_{h} \sigma^x(h)
	\label{eq:Hfree}
\end{equation}
where  the pseudo-spin operators $\sigma^x(h)$  are the kinetic energy
terms    defined     in    Table~\ref{tab:loops}        (see      also
appendix~\ref{sec:ZE}).     This  model    is   obtained  by   setting
$h_\alpha=1$  in   Eq.~\ref{eq:heffkag}.   We showed~\cite{msp02} that
Eq.~(\ref{eq:Hfree}) is completely  solvable  and is the prototype  of
RVB dimer liquid.  It  has a unique  ground-state (up to a topological
degeneracy) and  its elementary excitations are  pairs of gapped Ising
vortices   (visons).  Here, we    search   for  a  model   which,   as
Eq.~\ref{eq:Hfree}, is amenable  to an analytical treatment  and which
could  capture some essential features of  the  spin model.  The first
step in treating  the frustration inherent to  the Heisenberg model on
the kagome lattice   is to introduce  non trivial  {\em signs} in  the
dimer resonance loops.

\subsection{Definition of $\mu(h)$}

We choose to keep only the sign of the leading terms $h_\alpha$ of the
dimer overlap  expansion for the Heisenberg  model. This leads  to the
definition of an  operator  $\mu(h)$ at  each pseudospin location  $h$
(hexagon centers in the kagome case):
\begin{equation}
	\mu(h) = \sum_{\alpha=1}^{32}
	\epsilon(\alpha)
	\left|L_\alpha\right>\left<\bar{L}_\alpha\right|
	+
	{\rm h.c.}
\label{eq:mu_def}
\end{equation}
where
\begin{equation}
	\epsilon(\alpha)=(-1)^{{\rm Length}(\alpha)/2}
\end{equation}
and  ${\rm Length}(\alpha)$ is the length  of  the loop $\alpha$.  The
action of $\mu(h)$ only  differ from the pseudospin flip $\sigma^x(h)$
by  a sign: $\mu(h)\left|D\right>=\pm\sigma^x(h)\left|D\right>$.  This
sign depends on the  length  of the  admissible  loop at $h$  in state
$\left|D\right>$, as  indicated in Table~\ref{tab:loops}.   With  this
definition,    if   $\left|c\right>$ is  a  (non-orthogonalized) dimer
configuration, we have $\left<c\right| \mu(h) \left|c\right>\ge 0$ for
any hexagon $h$. This can be used  to define the signs matrix elements
of  $\mu$ independently  of the  orientation  of the  dimers.

\begin{figure}
	\begin{center}
	\includegraphics[width=3.5cm,height=3.3cm]{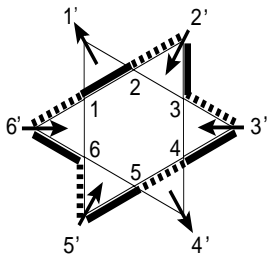}
	\end{center} \caption[99]{The  position of incoming  arrows on
	sites $1',2',\cdots$ defines the possible dimer move (full and
	dashed   links)  around the hexagon.    The   arrows at  sites
	$1,2,\cdots,6$ are  omitted for   clarity; they  take opposite
	directions  in the dimerizations pictured  by  full and dashed
	dimers  respectively.     The arrows at    $1',2',\cdots$  are
	unchanged during this dimer move.}\label{KagLoopWithArrows}
\end{figure}

As for $\sigma^x$, $\mu$ can be simply expressed in terms of the arrow
representation introduced  in     Sec.~\ref{sec:arrow}.      Let
$1,2,\cdots,6$ be the sites  of hexagon $h$ and $1',2',\cdots,6'$  the
sites   of   the      ``star''   surrounding   this   hexagon     (see
Fig.~\ref{KagLoopWithArrows}).    The length  of  the admissible  loop
$\alpha_0$  around  $h$  is related   to the  state of  the  arrows on
$1',2',\cdots,6'$ in the following way.  If  the arrow at site $i'$ is
pointing  toward $h$, it shares  a dimer with  a site $j$ belonging to
the hexagon  and $\alpha_0$  will  pass  through  $i'$  and $j$.   Let
$n_{\rm out}$ be the number of outgoing arrows.  From this it is clear
that      the     length        of    that       loop      will     be
${\rm{Length}}(\alpha_0)=12-n_{\rm{out}}$.  To summarize, the operator
$\mu(h)$ defined in  Eq.~\ref{eq:mu_def}  flips  the arrows of   sites
$1,2,\cdots,6$ and multiplies  the configuration by a  sign\footnote{
The length $L_{\rm  max}$ of longest loop around  a star is always even
in our  family of lattices  since $L_{\rm max}=2L_{\rm min}$. For this
reason Eq.~\ref{snout} is valid for any such lattices.}:
\begin{equation}
	\epsilon(h)=(-1)^{n_{\rm{out}}(h)/2}
	\label{snout}
\end{equation}

The operators $\mu$  can   be explicitly  written with the    help of
$\sigma^z$ and $\sigma^x$   operators.  Finding such  an expression is
not completely obvious  since the $\sigma^z$  are non local and depend
on  the reference   configuration whereas  the  $\mu$  are  local  and
independent of any  reference  state.   The expression is  derived  in
appendix~\ref{sec:Sigmaz2Mu}. 

\subsection{Commutation rules}

\subsubsection{Operator $\mu$}

\begin{figure}
	\begin{center}
	\resizebox{!}{2.6cm}{\includegraphics{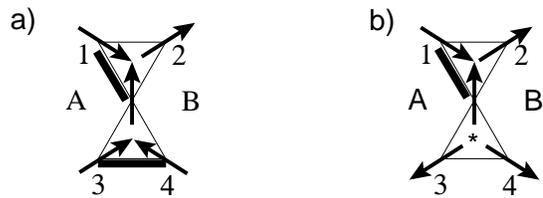}}
	\end{center} \caption[99]{Two possible states of
	a pair of triangles.}\label{TwoTriangles}
\end{figure}

The  $\mu$ operators have the  remarkable property of 1) anticommuting
when they operate   on  nearest-neighbor  hexagons but  2)   commuting
otherwise.    This  property is     not    obvious when  looking    at
Eq.~\ref{eq:mu_def} and is  most easily demonstrated  with the help of
the      arrow   representations.   
As mentioned previously, the effect of  $\mu(h)$ on an arrow configuration is
to flip the arrows around the hexagon $h$ and multiply it by a sign 
$\epsilon(h)=(-1)^{n_{out}(h)/2}$. Therefore, the action of $\mu(A)$ and $\mu(B)$ commute up to a sign. If the hexagons $A$ and $B$ are not adjacent,
the signs $\epsilon(A)$ and $\epsilon(B)$ are unaffected by the action
of $\mu(B)$ and $\mu(A)$ respectively, and
$\mu(A)$ and $\mu(B)$ commute.
For two neighboring hexagons $A$ and $B$, the action of $\mu(A)$
affects the sign $\epsilon(B)$ and conversely.
There are two types of arrow configuration shared by the 
neighboring hexagons $A$ and $B$, as shown in Fig. ~\ref{TwoTriangles}a 
and b. Both configurations have an odd number of outgoing arrows 
among the four external links which
will be flipped by the successive action of $\mu(A)$ and $\mu(B)$.
One of the signs $\epsilon(A)$ and $\epsilon(B)$ will therefore
be changed by the action of the neighboring $\mu$, but not the other.
This means that  upon acting with $\mu(A)$ and $\mu(B)$, the sign of the final configuration depend on the order we applied the two operators,
and we find that on any configuration $\mu(A)\mu(B)=-\mu(B)\mu(A)$.

\subsubsection{Operator $\mut$}

Another choice for the signs of  the tunneling amplitudes turns out to
be very useful. Consider the $\mut$ operators defined by
\begin{equation}
	\mut(h) = \sum_{\alpha=1}^{32}
	\tilde{\epsilon}(\alpha)
	\left|L_\alpha\right>\left<\bar{L}_\alpha\right|
	+
	{\rm h.c.}
\label{eq:mut_def}
\end{equation}
where  the  signs  $\tilde{\epsilon}(\alpha)$  are given  in  the last
column  of  Tab.~\ref{tab:loops}.     Contrary  to $\epsilon(\alpha)$,
$\tilde{\epsilon}(\alpha)$ counts the parity of the number of outgoing
arrows  only   on one-half  of  the sites   of  the  star: $1',3'$ and
$5'$. Using the  arrow representation and  similar arguments as above,
one can  show that  1) $\mut$   operators anticommute when  acting  on
nearest-neighbor hexagons 2) commute otherwise. Most interestingly the
$\mu$ and $\mut$ realize  two copies of  the same algebra that commute
with each other:
\begin{eqnarray}
	\forall h,h' \;\;\;\; \left[\mu(h),\mut(h')\right]=0
\end{eqnarray}

The $\mu$  (and  $\mut$) operators  have simple  commutation relations
with   the  pseudo-spin operators $\sigma^z$  introduced    by ZE.  By
definition       the   $\mu$    operators       can       be   written
$\mu(h)=\epsilon(h)\sigma^x(h)$    where    $\epsilon(h)$  is diagonal
operator  in the dimer basis.  Because  $\sigma^z$  operators are also
diagonal in that basis they commute  with any $\epsilon$ and we simply
have           $\mu(h)\sigma^z(h)=-\sigma^z(h)\mu(h)$              and
$\left[\mu(h),\sigma^z(h')\right]=0$ for $h\ne         h'$        (see
appendix~\ref{sec:ZE}   for the commutation  rules   of $\sigma^x$ and
$\sigma^z$ operators).  The  same  result holds for  the  $\mut$.   In
Ref.~\onlinecite{msp02}  we  identified $\sigma^z(h)$  as the operator
which creates (or destroy) an Ising vortex (vison)  on hexagon $h$ and
a non-zero expectation value $\left<\sigma^z\right>$ was interpreted as
signature of confinement and vison condensation.

\subsection{Hamiltonian}
The main results  of this paper  concern  the following quantum  dimer
model
\begin{equation}
	\mathcal{H}_\mu =-\sum_h\mu(h),
	\label{hmu}
\end{equation}
where  $\mu$ obey  the commutation  rules  described in  the previous
section.     We   studied  this    model    both    numerically    and
analytically. Numerically  we  diagonalized it on   systems up to  144
kagome     sites (48 hexagons).       The   results are  presented  in
section~\ref{sec:numerics}. The most striking  feature of the spectrum
is that all energy  levels have a  huge extensive degeneracy of  order
$2^{N_{ps}/2}$, where, in the kagome  case, $N_{ps}=N/3$ is the number
of hexagons.  The degeneracy of the spectrum
has its origin in  the  existence of the  set of  operators $\mut(h)$,
which     commute     with    the   $\mu$,      and    therefore  with
$\mathcal{H}_\mu$.  The  spectrum is   
organized in  $2^{N_c}$ sectors
labeled  by the   eigenvalues  of the   $N_c$   independent commuting
operators  which can be  constructed from the $\mut$, as explained in Sec.~\ref{sssec:nodegrep}.  The eigenvalues
are identical in all the $2^{N_c}$ sectors.
Another interesting feature is the existence of quantities which
commute both with $\mu$ and $\mut$, called in the following
integrals of motion. They are constructed from products 
of $\mu$ (or alternatively $\mut$) on straight lines drawn on 
the triangular lattice of pseudo-spins, as explained in 
Sec.~\ref{ssec:fermion}. The spectrum {\it depends}
on the values of these integrals of motion.

Using the operators  $\mut$, the dimer-dimer correlations in the model
can be shown (Sec.~\ref{sssec:correl}) to be short ranged.


\subsection{One-dimensional $\mu$-model}
\label{sec:1d}

\begin{figure}
	\begin{center}
	\includegraphics[width=7.5cm]{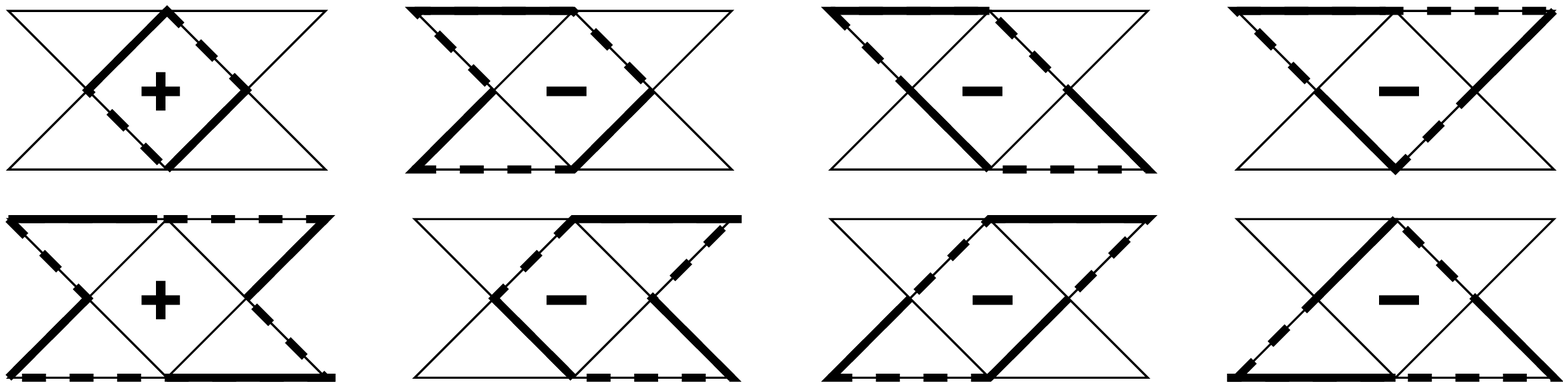}
	\end{center}

	\caption[99]{The eight possible  dimer  moves around  a square
	plaquette     of    the     frustrated    ladder    shown   in
	Fig.~\ref{chain}. The signs of the corresponding amplitudes in
	the $\mu$-model are indicated. }\label{fig:muc}

\end{figure}

Before  discussing the kagome case in more detail,  
it is interesting  to look at the
$\mu$-model     on     the    one-dimensional  lattice       displayed
Fig.~\ref{chain}. As on any lattice  made of corner-sharing triangles,
the $\mu$-model can be defined  there.  The different dimer moves (and
their       signs   $\epsilon(\alpha)$)    are   displayed          in
Fig.~\ref{fig:muc}. Because it  is one-dimensional, it is solvable and
we  will show that  its spectrum exactly maps to  the  spectrum of the
{\em transverse-field  Ising chain at    the critical field}  (quantum
critical point).

\subsubsection{Transverse-field Ising chain}

The  mapping to the transverse-field Ising chain can be realized 
through a representation of  the
algebra  of the $\mu$ by  some  pseudo-spin operators $\tau^x$  and
$\tau^z$ defined by:
\begin{eqnarray}	
	\tau^x_{2n}=\mu_{2n} &\;\;&
	\tau^x_{2n+1}=\mut_{2n} \nonumber \\
	\tau^z_{2n}= \prod_{l\le n} \mu_{2l-1} &\;\;&
	\tau^z_{2n+1}= \prod_{l\le n} \mut_{2l-1} 
	\label{eq:ising2mu}
\end{eqnarray}
Using the (anti)commutation relations of $\mu$ and $\mut$ it is
easy to check  that $\tau^x_i$ and  $\tau^z_i$ anticommute.  The $\mu$
(resp.  $\mut$) operators only involve the Ising pseudospins on
even (resp. odd) sites:
\begin{eqnarray}
	\mu_{2n}=\tau^x_{2n}  &\;\;&
	\mut_{2n}=\tau^x_{2n+1}	\nonumber \\ 
	\mu_{2n+1}=\tau^z_{2n} \tau^z_{2n+2} &\;\;&
	\mut_{2n+1}=\tau^z_{2n+1} \tau^z_{2n+3}
	\label{eq:mu2ising}
\end{eqnarray}
And the $\mu$ Hamiltonian on the chain is now simply
\begin{equation}
	\mathcal{H}_\mu =
		-\sum_n \tau^x_{2n}
		-\sum_n \tau^z_{2n} \tau^z_{2n+2}
\end{equation}
which we  recognize  to be  that  of  a ferromagnetic   Ising chain in
transverse  field at its  critical point~\cite{sachdev99}.  This model
can be  solved   by a  Jordan-Wigner transformation   (reduces to free
fermions). Because the Ising chain is at its critical point, the dimer
spectrum  is gapless and   support linearly dispersive excitations  at
small momentum.  In addition, $\left<\mu_i \mu_j \right>$ correlations
decay  algebraically  with distance.    The exponential  degeneracy of
$\mathcal{H}_\mu$ is now transparent: only one half of the Ising spins
(located  on  even   sites)    appear     to  be  coupled   by     the
Hamiltonian. However this  entropy has a  subtle origin: To  write the
the $\mu$-model {\em  only   in terms  of  the $N_{ps}/2$  ``coupled''
degrees of freedom} (in  order to get rid  of the entropy) one  has to
use operators  ($\tau^z_{2n}$) which  are  non-local for the  original
dimers (see  Eq.~\ref{eq:ising2mu}).  On  the  other hand,  there  are
local operators (the $\mut$ themselves)  that do not change the energy
and  which create localized  zero-energy  excitations.

We can make a comparison with another quasi one-dimensional model with
extensive degeneracy:  the spin-$\frac{1}{2}$ Heisenberg model defined
on a chain of  coupled  tetrahedra~\cite{mtm99}.  In that  model  some
Ising-like degrees of   freedom  $\chi=\pm1$ (spin chirality) do   not
appear in the Hamiltonian of  the low-energy sector  and the model has
an extensive zero-temperature entropy.  This situation seems analogous
to  the   $\mu$-model:  the   $\tau^z_{2n+1}$  play the   role  of the
$\chi$. However the important difference is that the $\tau_{2n+1}$ are
non-local in terms of the original dimers whereas the $\chi$ are local
in term of the original spins.

We  will see in  the next sections  that the $\mu$-model on the kagome
lattice  also has this important  property that  the ``coupled'' Ising
degrees of freedom are non-local in terms of dimers.  We are not aware
of any  other  interacting  quantum   model exhibiting such   kind  of
extensive degeneracies.  Their might be however  an analogy with other
systems  with  localized    excitations  such  as   in   Aharonov-Bohm
cages~\cite{vidal98} or  flat-band systems   in general.   There,  the
extensive degeneracy is due  to destructive interferences that prevent
excitations from hoping on the lattice and thus gaining kinetic energy
by delocalization.  This stresses  again the role  of the signs in the
$\mu$ operators.

\subsubsection{Order parameter}

The mapping  the to   the  transverse field-Ising  chain  suggests  to
introduce a different coupling for the $\mu$ on odd and even sites
\begin{eqnarray}
	\mathcal{H}(\Gamma)
		&=&-\Gamma\sum_n\mu_{2n}-\sum_n\mu_{2n+1} \\
		&=&-\Gamma\sum_n\tau^x_{2n}-\sum_n\tau^z_{2n} \tau^z_{2n+2}
	\label{eq:HG}
\end{eqnarray}
so that $\Gamma$ is the strength of the transverse field.  It is known
that                          the            ground-state          has
$\left<\tau^x_{2n}\right>\ne\left<\tau^z_{2n}\tau^z_{2n+2}\right>$
except  at the critical point  $\Gamma=1$.  In  other words, the $\mu$
have  different  expectation   values on  odd    and  even  sites  for
$\Gamma\ne1$.      This     state  ($\Gamma\ne1$)   has    long-ranged
$\left<\mu_i\mu_j\right>$  correlations   and  is a   crystal  in  the
$\left<\mu\right>$ variables   with   a  diverging   structure  factor
$S(\pi)=\sum_{n} (-1)^n\left<\mu_0\mu_n\right>$.  It is interesting to
remark that from  the Ising point  of view the natural order parameter
is the ``magnetization'' $\left<\tau^z_{2n}\right>$, which is non-zero
for  $\Gamma<1$ phase  but  vanishes  for $\Gamma\geq1$.  This   order
parameter  is     non-local    in   terms    of   the    dimers   (see
Eq.~\ref{eq:ising2mu}).  However   this  does    not mean    that  for
$\Gamma<1$ the  dimer system  spontaneously breaks  some  hidden Ising
symmetry. There is no such Ising symmetry in the dimer problem and the
spurious  $\mathbb{Z}_2$  redundancy   was introduced  in   the  Ising
representation: reversing  all the  $\tau^z$  gives in fact   the {\em
same}   physical   dimer  state,   as      it  can   be   seen    from
Eq.~\ref{eq:mu2ising}. In the dimer  language this is a consequence of
the fact that  reversing all the  arrows twice the  is proportional to
the identity.

\subsubsection{Heisenberg model on a frustrated ladder}

The spin-$\frac{1}{2}$ Heisenberg model on the three-spin ladder shown
in  Fig.~\ref{chain}    has    been studied  by  Waldtmann     {\it et
al.}~\cite{waldtmann00}.  They   considered a $J-J'$  model  where the
horizontal   coupling is $J$    and the  diagonal  one  (around square
plaquettes)  is    $J'$.    From    their numerical   results   (exact
diagonalizations  for $N\le30$ spins and DMRG  for $N\le120$ spins) it
appears that the system may  be critical in  the region $0.5  \lesssim
J'/J \lesssim 1.25$ (vanishing spin gap) and that a spin gap opens for
$J'/J \gtrsim 1.25$.  Quite interestingly  they showed that at  $J=J'$
the specific heat is quantitatively very similar to that of the kagome
antiferromagnet and exhibits a  low-temperature peak.  The finite-size
spectra  also show   a  large density of   singlet   states above  the
ground-state  (although  probably  not  exponential  in  $N$).   These
similarities suggest that the corner-sharing  triangle geometry is  an
important ingredient to generate  a large amount of low-energy singlet
excitations  and it would be  interesting to  investigate the possible
relation between this three-spin ladder and the $\mu$-model.

\subsubsection{Exact spectrum via fermion representation}

In  the mapping  to the transverse  field  Ising chain  have neglected
subtleties associated with boundary  conditions as well as constraints
on the physical  space  of dimer  coverings.   Indeed, the later   has
dimension $2^{N/3+1}$  whereas we used  a representation  of dimension
$2^{N_{ps}}=2^{N/3}$. We shall   now present the full solution of
the $\mu$-model on the chain, using fermionic variables.

Since we are  interested mainly in the  spectrum, we will  realize the
algebra of $\mu_j$ operators in a space which has the right 
dimension of the dimer
space, that  is $2^{N_{ps}+1}$. There exist an exact mapping between the 
fermionic states and the dimer states, but we do not insist on it here.   
We  suppose periodic boundary conditions and take $N_{ps}$ even.

Let us introduce a pair of creation/annihilation fermionic
operators $c_j,c_j^\dagger$ at each site $j=1,...,N_{ps}$.
They are equivalent to a pair of Majorana fermions,
$\gamma_j= (c_j^\dagger+c_j)$, $\tilde \gamma_j=
-i(c_j^\dagger-c_j)$, $\gamma_j^2=\tilde \gamma_j^2=1$.
We construct the operators
$\mu_j$ as follows
\begin{equation}
\label{eq:mu2gamma}
	\mu_j=i \gamma_j \gamma_{j+1}\;, \quad  \tilde \mu_j=
	i \tilde \gamma_j \tilde\gamma_{j+1}\;,
\end{equation}
so that
\begin{equation}
\label{eq:hamchaine}
	{\mathcal H}_\mu=-\sum_{j=1}^{N_{ps}}\mu_j=
	-i \sum_{j=1}^{N_{ps}}\gamma_j \gamma_{j+1}\;.
\end{equation}
It is straightforward to check that two operators $\mu_i$,
$\mu_j$
anticommute if they are neighbors and commute otherwise;
the same is valid for the operators $\tilde \mu_j$,
while $\mu_i$ and $\tilde \mu_j$ always commute.

We have not yet specified the  periodicity conditions on the operators
$\gamma_j$. Let us introduce two extra kinetic operators, $\mu(u)$ and
$\mu(d)$, which   move    the   dimers  around  the    two  edges  (or
$N_{ps}$-gones) of the chain   $$\mu(a)=\sum_{\alpha=1}^{2^{N_{ps}-1}}
\epsilon(\alpha)                          \left                      (
\left|L_\alpha\right>\left<\bar{L}_\alpha\right|   +    {\rm     h.c.}
\right)\;, \quad a=u,d $$ where  $\alpha$ runs over the $2^{N_{ps}-1}$
possible loops     of       even length   around    the      edge  and
$\epsilon(\alpha)=(-1)^{{\rm     Length}(\alpha)/2}$.   They  mutually
commute and  anticommute with all  $\mu_j$, $j=1,\ldots N_{ps}$. Their
product  $\mu(u)\mu(d)$  commute with the  Hamiltonian   and it is  an
integral of motion, taking the values  $\pm 1$. Other two integrals of
motion  are  the products of  $\mu_j$  on  the even,  respectively odd
sites,             $\mu_o=\mu_1\mu_3\ldots            \mu_{N_{ps}-1}$,
$\mu_e=\mu_2\mu_4\ldots \mu_{N_{ps}}$.  A careful analysis  shows that
$\mu_o\mu_e\mu(u)\mu(d)=(-1)^{3N_{ps}/2}$ on  any dimer state. This is
constraint on physical states shows that the  dimension of the Hilbert
space is  $2^{N_{ps}+1}$.   Suppose  now we  are   in the sector  with
$\mu(u)\mu(d)=1$.  Then, $\mu_o\mu_e=-(-1)^{N_{ps}/2}\mu_1\mu_2\ldots
\mu_{N_{ps}}$,    and     using    the    definition    from  equation
(\ref{eq:mu2gamma})   we  obtain  that $\gamma_1\gamma_{N_{ps}+1}=-1$,
which  implies  antiperiodic  boundary   conditions   on the  Majorana
fermions. In    the second sector,  $\mu(u)\mu(d)=-1$,   the  Majorana
fermions have periodic boundary conditions.

We have now all that we need to solve the model (\ref{eq:hamchaine}).
First, all the spectrum is
degenerate $2^{N_{ps}/2}$ times, since all the $N_{ps}$ Majorana fermions
$\tilde \gamma_j$ commute with the Hamiltonian.
Then, the Hamiltonian  \ref{eq:hamchaine} can be diagonalized after a Fourier
transformation,
\begin{equation}
\label{eq:hamchaine2}
	{\mathcal H}_\mu=2\sum_{k}\sin k\; \gamma_{-k} \gamma_{k}\;.
\end{equation}
where the sum is over momenta $k=2\pi n/N_{ps}$, with
$n$ being an integer in the periodic sector a
half-odd integer in the antiperiodic sector, $0\leq n<N_{ps}$.
The operators $\gamma_k$ with $0<k<\pi$ (resp. $-\pi<k<0)$ could be
interpreted as annihilation (resp. creation) operators. Note that, in order to ensure
the right commutation relations between creators and annihilators,
we defined the Fourier modes as $\gamma_k=1/\sqrt{2N_{ps}}\sum_n{\rm e}^{ikn}\gamma_n$. The zero
modes $k=0, \pi$ need separate treatment, but they
do not appear in the Hamiltonian. After normal ordering,
\begin{equation}
\label{eq:hamferm}
	{\mathcal H}_\mu=4\sum_{0<k<\pi}|\sin k|\  \gamma_{-k} \gamma_{k}
	-2\sum_{0<k<\pi}|\sin k|\;.
\end{equation}
The vacuum energy is easily calculated in the two sectors,
\begin{eqnarray}
E_0^{(-)}&=&-2\sum_{n=0}^{N_{ps}/2-1} \sin \frac{\pi(2 n+1)}{N_{ps}}
=-\frac{2}{\sin(\pi/N_{ps})}
\;,\cr
E_0^{(+)}&=&-2\sum_{n=1}^{N_{ps}/2-1} \sin \frac{2\pi n}{N_{ps}}
=-\frac{2}{\tan(\pi/N_{ps})}\;.
\end{eqnarray}
Since $E_0^{(-)}<E_0^{(+)}$, the ground state of the dimer problem has
energy $E_0=E_0^{(-)}=-2/\sin(\pi/N_{ps})$. In the thermodynamic limit
$E/N_{ps}\to-2/\pi$.   The first excited state  is at $E_0^{(+)}$; the
rest  of  the  spectrum  can  be constructed   by making particle-hole
excitations   over   the   two     fermionic   vacua,   according   to
Eq.~\ref{eq:hamferm}. The numerical  spectra, obtained in the
dimer representation, are in complete agreement
with   the    ones    constructed   from~\ref{eq:hamferm}.    In   the
thermodynamic limit the gap  vanishes and the excitation spectrum of
the $\gamma_k$ is linear at small momentum.

\subsection{Kagome case}
\label{sec:mukag}
\subsubsection{Degenerate representation}

As we did in the one-dimensional case,  it is natural to represent the
$\mu$ operators with Ising variables. The simplest representation uses
one Ising variable $\tau^z=\pm1$ at each hexagon:
\begin{eqnarray}
	\mu_{\bf r}=\tau^x_{\bf r} \prod_{i=1}^{3} \tau^z_{{\bf r}+{\bf e}_i}
	&\;\;&
	\mut_{\bf r}=\tau^x_{\bf r} \prod_{i=1}^{3} \tau^z_{{\bf r}-{\bf e}_i}
	\label{eq:degrep}
\end{eqnarray}
where the three unit vectors ${\bf e}_i$ are at 120 degrees and relate
a site to three of its neighbors on the triangular lattice. It is easy
to  check  that this  representation   indeed realizes the  $\mu$ (and
$\mut$) algebra.   One can in  fact express the $\mu$  in  terms of ZE
pseudospin operators $\sigma^z$  and    find similar  (although   more
complicated)  relations (see appendix~\ref{sec:Sigmaz2Mu}). Notice  in
particular that  $\tau^z$ anticommutes with   $\mu$ and $\mut$  on the
same hexagon   but  commutes with all    the others, exactly    as the
$\sigma^z$ do.  This   representation has (approximately) the  correct
size of the Hilbert space ($\sim 2^{N_{ps}}$) but it does not show how
many Ising variables decouple in this model, that  is how large is the
degeneracy (entropy) in this model.

\subsubsection{Three-sublattice representation}

One can use a different  representation for the  $\mu$'s on a  smaller
subspace, therefore   removing  part of  the  degeneracy.   Consider a
decomposition of the  triangular lattice into {\em three}  sublattices
$A,B$ and $C$.  All the   $\mut(c)$ with $c\in{C}$ commute with   each
other  (as well as  with all the  $\mu$).  They  can be simultaneously
diagonalized   so   that     we    can   consider    an     eigenstate
$\left|\psi_0\right>$ of these $\mut(c)$:
\begin{equation}
	\mut(c)\left|\psi_0\right>=\eta(c)\left|\psi_0\right>
	\;\;\;\; \eta(c)=\pm1
	\label{eq:eta}
\end{equation}
In addition we may define some operators $s(h)$ by:
\begin{equation}
	s(h)=\mu(h)\mut(h)
\end{equation}
They are diagonal in  the dimer basis and commute  with each other. We
can   project    $\left|\psi_0\right>$   onto   the   subspace   where
$s(c)=\eta(c)$ for any site $c\in C$:
\begin{eqnarray}
	\left|\psi_1\right>&=&
	\prod_{c\in C}\left(\frac{1+\eta(c)s(c)}{2}\right)
	\left|\psi_0\right>
\end{eqnarray}
which, by Eq.~\ref{eq:eta}, is simply
\begin{eqnarray}
	\left|\psi_1\right>&=& \prod_{c\in C}\left(\frac{1+\mu(c)}{2}\right)
	\left|\psi_0\right>
\end{eqnarray}

Now   we   consider the    states generated  by    the action   of the
$\mu(a{\in}A)$ and $\mu(b{\in}B)$ on $\left|\psi_1\right>$. A basis can
be    labeled  by $2N_{ps}/3$   Ising    variables $\tau^z_a=\pm1$ and
$\tau^z_b=\pm1$ as follows:
\begin{equation}
	\left| \tau^z_a,\tau^z_b \right>
	=
	\left[\prod_{a\in A} \mu(a)^{\frac{1}{2}(1+\tau^z_a)}\right] \;
	\left[\prod_{b\in B} \mu(b)^{\frac{1}{2}(1+\tau^z_b)}\right]
	\left|\psi_1\right>
	\label{eq:3sub}
\end{equation}
As we will now show, this  subspace is stable  under the action of any
$\mu$.  This  is obvious concerning   the operators $\mu$  located  on
sublattices $A$  and $B$. On  these sites we may define (non-diagonal)
$\tau^x$ operators   which reverse   the   value of $\tau^z$   at   the
corresponding site.  With this definition we have
\begin{eqnarray}
	\mu(a)&=&\tau^x_a \\
	\mu(b)&=&\tau^x_b \tau^z_a \tau^z_{a'} \tau^z_{a''} 
\end{eqnarray}
The $\tau^z_a\tau^z_{a'}\tau^z_{a''}  $ comes from the anticommutation
of $\mu(b)$  with its three neighbors  ($a$, $a'$ and $a''$) belonging
to $A$ when acting on a state like Eq.~\ref{eq:3sub}.  Now we act with
a $\mu(c)$ on $\left| \tau^z_a,\tau^z_b\right>$.  Upon moving $\mu(c)$
to the right through the $\mu(a)$ and $\mu(b)$,  the state picks a sign
$\tau^z_a\tau^z_{a'}\tau^z_{a''}\tau^z_b\tau^z_{b'}\tau^z_{b''}$ where
$a,a',\cdots,b''$ are the six neighbors of  $c$.  Then we use the fact
that  $\mu(c)\left|\psi_1\right>=\left|\psi_1\right>$  so   that    we
finally get the following representation:
\begin{eqnarray}
	\mu(a)&=&\tau^x_a \nonumber \\
	\mu(b)&=&\tau^x_b \tau^z_a \tau^z_{a'} \tau^z_{a''} \nonumber\\
	\mu(c)&=&\tau^z_a \tau^z_{a'} \tau^z_{a''}\tau^z_b \tau^z_{b'} \tau^z_{b''}
	\label{eq:muabc}
\end{eqnarray}
This representation is independent of the $\eta(c)$. As a consequence,
the spectrum  of any  Hamiltonian made  out  of $\mu$  operators (like
$\mathcal{H}_\mu$) will  be at  least $2^{N_{ps}/3}$ fold  degenerate.
As we show  later that  this degeneracy  is in fact  much larger $\sim
2^{N_{ps}/2}$ but this already exhibits  an extensive residual entropy
at zero temperature.

\subsubsection{Mean-field approximation}
\label{sssec:abc}
The representation of Eq.~\ref{eq:muabc} suggests a simple variational
(or  mean-field) approximation in  which the  system   is in a  tensor
product               of       single-site              wave-functions
$\left|\Psi\right>=\bigotimes_1^{N_{ps}/3}\left(\left|\Psi_A\right>\otimes\left|\Psi_B\right>\right)$. All
the sites $a\in A$ are in the same state as well as all the $b\in B$:
\begin{eqnarray}
	\left|\Psi_A\right>&=&
	\cos(\theta/2)\left|\uparrow\right>+\sin(\theta/2)\left|\downarrow\right>
	\\
	\left|\Psi_B\right>&=&
	\cos(\phi/2)\left|\uparrow\right>+\sin(\phi/2)\left|\downarrow\right>
\end{eqnarray}
We have two variational parameters $\theta$ and $\phi$
and the expectation value of the energy per hexagon is
\begin{eqnarray}
	\left<\mu_a\right>&=& \sin(\theta) \nonumber \\
	\left<\mu_b\right>&=& \cos(\theta)^3\sin(\phi)  \label{eq:mfmuabc}\\	
	\left<\mu_c\right>&=& \cos(\theta)^3\cos(\phi)^3 \nonumber \\
	E/N_{ps}&=-&\frac{1}{3}\left(
		\left<\mu_a\right>+\left<\mu_b\right>+\left<\mu_c\right>
		\right) \label{eq:mfe}
\end{eqnarray}
Minimizing $E$ we get
\begin{eqnarray}
	\left<\mu_a\right>\simeq0.2979 \;;\;
	\left<\mu_b\right>&\simeq&0.3104 \;;\;
	\left<\mu_c\right>\simeq0.7091
	\label{eq:<muabc>}
	\\
	E/N_{ps}&=&-0.4391 \label{eq:Eabc}
\end{eqnarray}
Because
$\left<\mu_a\right>\neq\left<\mu_b\right>\ne\left<\mu_c\right>$   this
state breaks the translation  symmetry.  It has some crystalline order
with respect to  the  $\left<\mu\right>$ variables.  Notice,  however,
that such a state is {\em  not a dimer crystal} (see Eq.~\ref{eq:nn}),
since it can be chosen to have  zero dimer-dimer correlations beyond a
few lattice spacings.  From the numerical diagonalizations we estimate
the exact ground-state energy  to be $E/N_{ps}\simeq-0.44\pm0.02$ (see
section~\ref{sec:numerics} and  Fig.~\ref{fig:gap}), which agrees with
the variational  result  within error   bars.   It is  interesting  to
compare these energies with the energy that one get with a single-site
wave-function derived from the translation invariant representation of
Eq.~\ref{eq:degrep}.   The     later          approximation      gives
$E/N_{ps}=-0.3248$,   which  is   significantly  higher.   Two  others
mean-field states can  be considered from Eq.~\ref{eq:degrep}.  On can
use three    different    single-spin  states   $\left|\Psi_A\right>$,
$\left|\Psi_B\right>$     and    $\left|\Psi_C\right>$    on     three
sublattices. Minimizing the energy  with respect to the  three related
angles we get   $E/N_{ps}=-1/3$.  The corresponding  variational state
simply                 has                     $\left<\mu_a\right>=1$,
$\left<\mu_b\right>=\left<\mu_c\right>=0$.   Enlarging   the unit cell
does not help to  lower  the energy.   Indeed, using four  sublattices
leads to an even worse energy ($E/N_{ps}=-1/4$, $\left<\mu_a\right>=1$
and $\left<\mu_{b,c,d}\right>=0$).     The fact  that   the degenerate
representation  gives rather bad  energies  can be  explained from the
fact in  such states   the  Ising  degrees  of freedom  ($\tau^x$   or
$\tau^z$)    are    completely  uncorrelated  on    different   sites:
$\left<\tau_i\tau_j\right>=\left<\tau_i\right>\left<\tau_j\right>$. The
situation is different in a representation like Eq.~\ref{eq:muabc}. In
that case non-trivial  nearest-neighbor correlations are present  even
in simple tensor-product states such as the one we considered.

Although the good variational  energy given Eq.~\ref{eq:Eabc} does not
prove that  the  system  indeed  spontaneously  break the  translation
symmetry, it  indicates that the $\mu$-model on  the kagome lattice is
not very far  from such a  phase\footnote{It is not surprising to find
such a three-sublattice  order in this variational approximation since
the translation      invariance     is    explicitly    broken      by
Eq.~\ref{eq:muabc}.}.       The    numerical   results   presented  in
section~\ref{sec:numerics} indeed show   that,   at least   at   short
distances,   $<\mu(x)\mu(y)>$  correlations match the three-sublattice
pattern.

\subsubsection{Three-sublattice $\mu$-model}
\begin{figure}
	\begin{center}      \includegraphics[width=3.5cm]{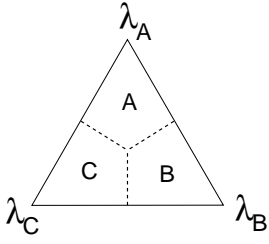}
	\end{center}  \caption[99]{Mean-field  phase diagram  of   the
	$\mu$-model   on  the kagome   lattice  with  three  different
	couplings  $\lambda_A$, $\lambda_B$  and $\lambda_C$ on  three
	sublattices.  Because the mean-field approximation breaks the
	symmetry which exchanges   $A$,  $B$ anc $C$,  the  transition
	lines (dashed) do not precisely match the symmetry axis.  Such
	symmetry should hold in the real  system and was restored here
	for clarity. Transitions are first order.}\label{fig:mfmuabc}
\end{figure}

As  for  the  one-dimensional  model, we   can  generalize the  kagome
$\mu$-model by letting the  couplings be different on sublattices $A$,
$B$ and $C$.
\begin{equation}
	\mathcal{H}
	=-\lambda_A\sum_{a\in A} \mu_a
	-\lambda_B\sum_{b\in B} \mu_b-\lambda_C\sum_{c\in C}\mu_c
\end{equation}
and         we  focus   on   $\lambda_A+\lambda_B+\lambda_C=1$     and
$\lambda_A,\lambda_B,\lambda_C\geq0$.  We determined the  ground-state
of  the    model     within    the  mean-field   approximation      of
Eqs.~\ref{eq:mfmuabc}-\ref{eq:mfe}.  The result is schematically shown
Fig.~\ref{fig:mfmuabc}  (a  qualitatively similar  phase   diagram  is
obtained with the degenerate representation).  We obtain three phases.
When   $\lambda_A$  dominates  the   ground-state  has $\theta=\pi/2$,
$\left<\tau^z_a\right>=0$    and $\left<\tau^z_b\right>\ne0$.     When
$\lambda_B$      is    the     strongest   we   have     $\phi=\pi/2$,
$\left<\tau^z_a\right>\ne0$ and  $\left<\tau^z_b\right>=0$.  And close
to  the   $\lambda_C$ point  we  have  $\left<\tau^z_a\right>\ne0$ and
$\left<\tau^z_b\right>\ne0$.  Along the  transition lines (dashed line
in Fig.~\ref{fig:mfmuabc})  an expectation value $\left<\tau^z\right>$
jumps from 0 to a finite value so that the transitions are first-order
in this approximation.

The  mean-field  prediction  for  the topology   of the phase  diagram
appears to  be quite  plausible.   If there are  indeed three  phases,
then,    by      symmetry,       the      point       of      interest
$\lambda_A=\lambda_B=\lambda_C$   is  the   ending  point   of   three
transition lines.  It is not clear  whether the discontinuous character
of these transitions is an artifact of the mean-field approximation or
a robust  property.  If the transitions are  really first  order, then
the  $\mu$-model  spontaneously breaks the  translation invariance and
realizes a  crystal  in the $\mu$   variables.   It may also   be that
$\lambda_A=\lambda_B=\lambda_C$ is a critical point.  Although we have
no definite  conclusion on this issue,  some  of the numerical results
(susceptibility)    presented in Sec.~\ref{sec:numerics} suggests    a
critical point.

\subsubsection{Non-degenerate representation}
\label{sssec:nodegrep}

\begin{figure}
	\begin{center}
	\resizebox{6cm}{!}{\includegraphics{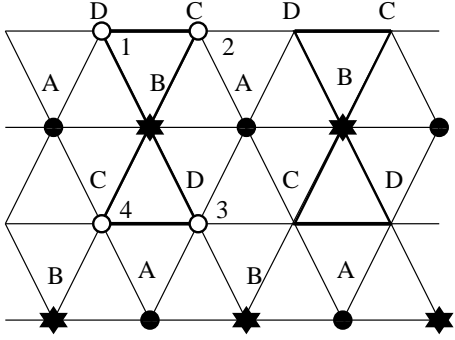}}  \end{center}
	\caption[99]{Dividing the triangular lattice in four sublattices
        $A,B,C$ and $D$. The open dots represent the ``spins''
        entering to a bow-tie operator centered on the sublattice $B$.
	The black dots and the stars correspond to the positions of spins and
	centers of the bow-tie operators respectively, forming a set of mutually 
	commuting operators.}\label{sabliers}
\end{figure}

The  constructions above   use at  most  $N_{ps}/3$ commuting   $\mut$
operators  but  this is   not he  maximum  number  $N_c$  of  mutually
commuting operators made out of the $\mut$.  In  fact there are of the
order of $N_c\simeq N_{ps}/2$  Ising degrees of freedom that decouple,
as it can be seen from the  following argument.  Divide the lattice in
{\em four} sublattices  $A,B,C,D$ as shown in the Fig.~\ref{sabliers}.
The  ``spins'' $\mut(h)$ on the  sublattice $A$  mutually commute.  In
addition,      one   can     consider  the   ``bow-tie''     operators
$\tilde{T}=\mut(1)\mut(2)\mut(3)\mut(4)$, centered on sites of the $B$
sublattice  and  involving neighboring  ``spins''of   the $C$ and  $D$
sublattices.  These $N_{ps}/4$  bow-tie operators commute mutually and
with the  operators $\mut(h)$ from  the sublattice $A$, which gives us
an  ensemble of  $N_{ps}/2$ commuting operators\footnote{They  are not
all independent on a finite  system with periodic boundary conditions,
in  the  sense that some products  on  closed straight lines reduce to
products of the integrals of  motion (monomials in the $\mu$ operators
- the $I_k$ defined below).  The number of dependent operators goes as
the number of   lines $M\;\sim\sqrt{N_{ps}}$, so  in the thermodynamic
limit they represent   negligible  fraction of   the total number   of
commuting operators.  The      degeneracy of the    spectrum  will  be
$2^{N_c}$, of the order of $2^{N_{ps}/2}$}.
These conserved  quantities ($\mut$ and  $\tilde{T}$) can be used
to eliminate de degrees of freedom on two sublattices in a similar way
as  we did for the  three-sublattice representation. However, now, one
cannot avoid obtaining a non-local representation of the $\mu$ algebra
in terms  of the spins   living on the sublattices   $C$ and $D$.  The
procedure is briefly explained in Appendix.~\ref{sec:red}.

\subsubsection{Arrow and dimer correlations}
\label{sssec:correl}

Using  the $\mut$ and $\tilde{T}$ introduced   above, we can show that
arrows and dimers are completely uncorrelated in any state provided it
is an eigenvector of the  $\mut(a)$ and $\tilde{T}(b)$ where $a\in{A}$
and $b\in{B}$ are hexagons in the vicinity of the arrows. The proof is
given below.

Let us  consider a triangle $(1,2,3)$  on the kagome lattice. On these
three sites we may define an arrow operator  $a_i$ whose value is 1 if
the corresponding arrow points toward the interior of the triangle and
0 otherwise.    With  this definition   the   dimer occupation  number
$n_{12}$ on bond $(12)$ is $n_{12}=a_1a_2$. Now we assume that we have
three operators $\hat{O}_i$ ($i=1,2,3$) which satisfy:
\begin{eqnarray}
	\forall i\;\;\;\;&& \hat{O}_i a_i=a_i\hat{O}_i \label{eq:aini}\\
	\forall i\ne j \;\;\;&& \hat{O}_i a_j= (1-a_j) \hat{O}_i \label{eq:ainj} \\
	&&\hat{O}_i \left|0\right>=\pm \left|0\right>
	\label{eq:Of}
\end{eqnarray}
These  relations just mean that $\hat{O}_1$  {\em flips} the arrows on
sites 2 and 3 but dot not touch the arrow 1, etc.  $\left|0\right>$ is
a ground-state  of  the model which   is also an  eigenvectors for  the
$\hat{O}_i$.     Notice that if    $\hat{O}_1$  and $\hat{O}_2$ exist,
$\hat{O}_3=\hat{O}_1  \hat{O}_2$ is  a   valid  choice.  Under   these
conditions, we show  that the  $n_i$  are uncorrelated  with any other
bond which is unaffected by the three $\hat{O}_i$.  As we will explain,
the  $\hat{O}_i$ will be realized  as local combinations of $\mut$ and
$\tilde{T}$. From Eqs.~\ref{eq:aini} and \ref{eq:ainj} it is simple to
check that
\begin{eqnarray}
	n_{12}\hat{O}_3 = \hat{O}_3 \left(1-n_{12}-n_{23}-n_{31}\right)
	\label{eq:nO}
\end{eqnarray}
(plus cyclic permutations).  Now let $X$  be any operator that commute
with the three  $\hat{O}_i$ (later we will  chose  $X=1$ or $X=n_{kl}$
where  $(kl)$ is  a    remote   bond).  Using  Eqs.~\ref{eq:Of}    and
\ref{eq:nO},   the   correlation  $\left<X\;n_{ij}\right>$   becomes:
\begin{eqnarray}
	\left< Xn_{12} \right>
		&=&\left<0\right|X n_{12}(\hat{O}_3)^2\left|0\right> \\
		&=&\left<0\right|
			\hat{O}_3X \left(1-n_{12}-n_{23}-n_{31}\right)\hat{O}_3
		\left|0\right> \\
		&=& \left< X \right>
		-\left< X n_{12}\right>
		-\left< X n_{23}\right>
		-\left< X n_{31}\right>
\end{eqnarray}
Using the relations obtained  by  cyclic permutations and solving  the
three linear equations we find:
\begin{eqnarray}
	\left< X n_{ij} \right>=\frac{1}{4} \left< X \right>
\end{eqnarray}
Using $X=1$ we  get that the  dimer density is $\frac{1}{4}$ and using
$X=n_{kl}$ (a remote link) we find
\begin{eqnarray}
	\left< n_{kl} n_{ij} \right>=\frac{1}{16}= \left< n\right>^2	
	\label{eq:nn}
\end{eqnarray}
In order to complete the demonstration we still  have to show that the
operators $\hat{O}_i$ can  be constructed for  any triangle.  Let  the
state $\left|0\right>$ be an eigenstate  of the $\mut(a{\in{A}})$  and
$\tilde{T}(b{\in{B}})$ described  in  the previous section.  Depending
on  its position relative  to the  four sublattices  $A$, $B$, $C$ and
$D$, a triangle $(123)$   (on kagome) will demand  slightly  different
constructions for its $\hat{O}_i$ operators.  For brevity we will only
consider the case of a  triangle located between an hexagon  $a\in{A}$
and an hexagon $b\in{B}$.   It  generalizes straightforwardly  to  the
other  cases. Let $i=1$ be the  common site of  hexagons  $a$ and $b$,
$i=2\in{a}$ and $i=3\in{b}$.   It  can be  checked  that the following
choice satisfies Eqs.~\ref{eq:aini} and \ref{eq:ainj} :
\begin{eqnarray}
	\hat{O}_1&=& \tilde{T}(b)\nonumber \\
	\hat{O}_2&=&  \mut(a)\tilde{T}(b) \\
	\hat{O}_3&=& \mut(a) \nonumber \\
\end{eqnarray}

The result given  Eq.~\ref{eq:nn} shows that  dimer-dimer correlations
are extremely short-ranged\footnote{For  Eq.~\ref{eq:nn} to hold, the
two bonds  must be at  a large enough distance  so that the $\mut$ and
$\tilde{T}$ involved in  the $\hat{O}_i$ of one  triangle do not touch
the second triangle.  This condition is automatically satisfied beyond
6 kagome-lattice spacings.}   in a ground-state like $\left|0\right>$,
it    is   a  dimer    liquid.   The   same   result   was    found in
Ref.~\onlinecite{msp02} in a  gapped  dimer model.  Here,  because the
ground-state manifold has a  huge  dimension, it  is likely  that some
perturbations are able to select  (out of the ground-state manifold of
$\mathcal{H}_\mu$) states  with some dimer order. Even  in such a case
we expect dimer-dimer correlations to be very weak  in the vicinity of
$\mathcal{H}_\mu$.

\subsubsection{Fermionic representation and integrals of motion}
\label{ssec:fermion}

A version of  the arrow representation  can be given in terms of fermions. 
The advantage of such a formulation is that signs are naturally
included. Unlike
the one-dimensional case, such  the fermionic representation cannot be
used   anymore to solve the  model,  since the Hamiltonian (\ref{hmu})
cannot be reduced to  a quadratic form  in  fermions.  However, it  is
still  useful in gaining some insight about  the model, for example  it  helps
understanding the extensive degeneracy of the spectrum.  In particular
it  provides  an argument which  shows   that the degeneracy is
$\sim 2^{N_{ps}/2}$  {\em not only on   the kagome lattice  but on any
lattice made of  corner-sharing  triangles}.   
Another advantage of  this formulation is to understand the origin
of the integrals of motions associated to products of $\mu$ along
straight lines.

We  associate  to each  vertex   of  the kagome  lattice a   fermionic
occupation number $0$ or $1$. For a given dimer configuration, the corresponding
fermion configuration is given by the following rules:  a defect triangle
has either occupation numbers $111$ or $000$. In the other triangles,
there is one dimer connecting the sites with equal fermion numbers,
$00$ or $11$.  There is a constraint on the parity of number of fermions
on each triangle, alternating on adjacent triangles, for example, triangles pointing  to the right in fig.~\ref{arrow} 
have odd  numbers of  fermions,  and that  pointing  to the left have  even
number of fermions. It is not difficult to see that the arrow variables
and the occupation number variables are essentially the same and that the constraints on them are of the same nature.
%
In particular, the counting  of the degrees of  freedom is similar for
arrows and fermions, the constraint  for each triangle eliminating one
spurious degrees of freedom.

On each kagome site there is a pair  of  creation/annihilation fermionic  
operators $c_j,c_j^\dagger$. As explained above, the dimer
space is equivalent to the fermion Fock  space, with constraints on
the occupation number on each triangle.

As in the case of the chain, we  transform  the fermions into a   pair of Majorana fermions,
$\gamma_j= (c_j^\dagger+c_j)$, $\tilde
\gamma_j=-i(c_j^\dagger-c_j)$. The algebra  of operators  $\mu(h)$ can
be realized  now    by using  only operators    $\gamma_j$  $$\mu(h)=i
\prod_j^{\rightarrow}  \gamma_j\;,$$  where the  arrow  means that the
product is oriented  (for example, it starts  at the leftmost  site of
the hexagon   and   runs clockwise).   With  this  representation  the
$\mu$-model contains 6  fermionic creation or annihilation  operators.
Since   two adjacent  hexagons have   one kagome site  in common,  the
associated $\mu$ operators anticommute. Distant $\mu$'s commute, since
they are constructed from even number  of fermions, and each operators
$\mu$ squares to $1$.  Similarly, one can construct $$\tilde \mu(h)=-i
\prod_j^{\leftarrow} \tilde \gamma_j\;,$$ with  the product running in
the opposite direction to that of $\mu$. These operators obey the same
algebra as $\mu(h)$ and commute with the whole set of operators $\mu$.
The  symmetric role  played by  the  operators $\gamma_j$ and  $\tilde
\gamma_j$ suggests that  one half  of the degrees  of freedom  are not
affected by the action of  the Hamiltonian Eq.~\ref{hmu} and therefore
that the degeneracy of the spectrum is  of  the order of $2^{N_{ps}/2}$.

Let us  note  that the  operators $\mu(h)$   and $\mut(h)$ leave   the
constraints on the occupation  numbers invariant, since they change by
two the occupation number  on each triangle. More  generally, products
of $\gamma$ on loops which visit each triangle an even number of times
also   leave  the constraints   invariant  and their   action could be
translated in  the dimer language.   The  most general operator  which
preserves the constraint  (dimer   space)  can be   constructed   from
products of $\gamma$ and   $\tilde  \gamma$  on loops  visiting   each
triangle an even number of times.   These operators are the equivalent
of Wilson loops in a gauge  theory.  Triangles where the constraint is
not obeyed can be    constructed by action   of strings  of   $\gamma$
operators (they could be  useful to describe dimer configurations with
defects, and therefore to  introduce holons or spinons).   And lastly,
the vison creation operator is  naturally constructed in this language
as the product of $\gamma_i\tilde\gamma_i$ on a string.

%

{\it  Integrals of motion. }
Let us analyze the case of closed systems
with the topology of the torus, which is the geometry we used for the
numerical diagonalization. The lattice on which 
the operators $\mu$ live is triangular and it is made by the centers
of hexagons of the kagome lattice.
Call $m_1$, $m_2$, $m_3$ the number of closed lines in each 
elementary direction on the lattice, 
having length $n_1$, $n_2$, $n_3$ respectively\footnote{We do not
consider the special case when one of the lines has length 2.}, 
so that $N_{ps}=m_1n_1=
m_2n_2=m_3n_3$. We denote by $M$ the total number of such lines,
$M=m_1+m_2+m_3$.
We label by $L_k$, $k=1,\ldots , M$ the closed straight lines
on the triangular lattice, and $l_{k}$ and $l_{k+1}$
the lines on the corresponding kagome lattice
bordering the line $L_k$ (with obvious periodicity conditions). 
Associated to each kagome line $l_k$, we can define the
following operators
	$$\Gamma_k=\prod_{j\in l_k}^{\rightarrow} \gamma_j\;,\qquad
	\tilde \Gamma_k=\prod_{j\in l_k}^{\leftarrow} \tilde \gamma_j\;,$$
where the products run over the sites $j$ of the kagome line $l_k$,
with some ordering indicated  by the arrow.
Since every line $l_k$ visits any hexagon 0 or 2 times,
the operators $\Gamma_k$ and $\tilde \Gamma_k$
commute with all the operators $\mu(h)$ and $\tilde \mu(h)$.
From the correspondence with the
dimer states, we know that  $\Gamma_k$ and $\tilde \Gamma_k$
correspond to dimer moves on non-trivial loops
around the torus, so they change the topological
sector.\footnote{Two operators $\Gamma_{k}$ and $\Gamma_r$
anticommute if they correspond to two independent
cycles, so we can choose $\Gamma_{k}$ and $\tilde \Gamma_r$ to
generate the four topological sectors on the torus.}

In each topological sector we can construct the following integrals of motion,
associated to the closed straight lines $L_k$ on the triangular lattice
	\begin{eqnarray*}
	I_k&=&(-i)^n \Gamma_k^{\rightarrow} 
	\Gamma_{k+1}^{\leftarrow}=\prod_{h\in L_k}^{\rightarrow} 
	\mu(h)\;,\\
	\tilde I_k&=&(-i)^n \tilde \Gamma_{k+1}^{\rightarrow}
	\tilde  \Gamma_k^{\leftarrow}=
	\prod_{h\in L_k}^{\leftarrow} \tilde \mu(h)\;
	\end{eqnarray*}
where $n$ denotes the number of hexagons on $L_k$ and the arrows
denote the ordering in $\Gamma_k, \tilde \Gamma_k$. The
the third member of both equalities is invariant by circular permutation
of the sites on the line $L_k$.

The two sets of integrals of motion $\{ I_k\},\ \{ \tilde I_k\}$ 
are not independent.
To check this, we use the commuting variables
	$$s(h)=\mu(h) \tilde \mu(h)\;,$$
measuring the parity number of fermions
around the hexagon $h$,
$$s(h)=\prod_j \gamma_j \tilde \gamma_j =(-i)^6 \prod_j (1-2n_j)\;.$$
Then,
$$I_k \tilde I_k=(-1)^n \prod_{h\in L_k} s(h) =(-1)^n\;,$$
where the last equality is a consequence of the 
constraint of fermions number around the two types of kagome 
triangles.
Moreover, integrals of motion corresponding to 
the lines $L_k$ with the same orientation on the lattice
are not all independent, 
	\begin{equation}
	\prod_{k=1}^{m_1} I_k=(-i)^{m_1n_1}=(-i)^{N_{ps}}\;,
	\end{equation}
an similarly for the other two orientations.
Such a constraint on physical states is not unexpected, since the product of
the lines with some orientation contains all the operators
$\mu(h)$ exactly once, that is all the operators $\gamma_i$
twice, so it has to be proportional to the identity.
 
In  conclusion,  there  are at   most $M-3$ independent  integrals  of
motion, where $M$ is the number of different closed straight lines one
can draw on  the  triangular lattice wrapped   on the torus  (in  some
cases, the $M-3$ lines are not independent and  some of the quantities
$I_k$ can be written as products of the others).

{\it  Basis for the Hilbert space. }
The commuting quantities $s(h)$ and $I_k$ can be used to label the states
in the Hilbert space. Let us first check that we obtain the right dimension.
Due to the $M-2$ independent  constraints $\prod_{h\in L_k} s(h)=1$, only
$N_{ps}-(M-2)$ of the operators $s(h)$ are independent.
Both type of operators $s(h)$ and $I_k$ can take only two
values, $\pm 1$, or $\pm i$ for $I_k$ on a line of odd length.
Taking into account the topological degeneracy, the
number of states in the Hilbert state is $4\times 2^{N_{ps}-(M-2)}\times
2^{M-3}=2^{N_{ps}+1}$~, which is the right dimension of the dimer space
(in the case when there are extra relations between the lines $L_k$,
the number of independent integrals of motion and the number of
constraints on $s(h)$ are diminished by the same number).


\section{Exact diagonalizations of the $\mu$-model}
\label{sec:numerics}

\begin{figure}
	\begin{center}
	\includegraphics[width=5cm,bb=18 144 310 719]{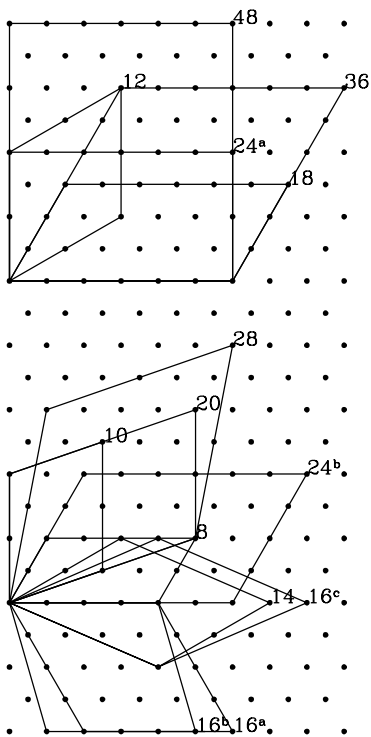}
	\end{center}
	\caption[99]{Finite-size lattices  used     for  the numerical
	diagonalizations of the $\mu$-model. Each dot is an hexagon of
	the kagome lattice.  $N_{ps}=\frac{1}{3}N$  is indicated.  The
	lattices on  the top are  compatible  with a  three sublattice
	structure  whereas the  others (bottom) are  not. All lattices
	except   $N_{ps}=10,14,16^b,16^c$  are   compatible with   the
	four-sublattice structure. }\label{fig:latt}
\end{figure}

Because  the non-diagonal matrix elements  of the  $\mu$ operator have
different signs, $\mathcal{H}_\mu$ is  not appropriate for large-scale
Monte-Carlo simulations  and,  instead,  we performed some   numerical
diagonalizations  of   the   Hamiltonian.      For  systems  up     to
$N_{\rm{hex}}=N_{ps}=16$   hexagons ($N=48$  sites)  we diagonalize it
directly   in  the  basis of   dimer   coverings  (which  dimension is
$2^{N_{ps}+1}$) by using all  lattice symmetries.  For  larger systems
($N_{ps}=20$,  24,   28, 36)  we  diagonalize  the   Hamiltonian in  a
representation where the extensive  degeneracy (due to the $N_c$ Ising
quantities that  commute with  every  $\mu_i$) have  been removed (see
appendix~\ref{sec:red}).  The diagonalization is performed  separately
for each  sector defined  by the conserved   quantities $I$.   For the
largest system ($N_{ps}=48$, $N=144$) we   use a Lanczos algorithm  in
this non-degenerate  representation  to obtain the  first energies and
wave-functions. Thanks to these numerous symmetries the largest vector
size is  only $\sim   10^6$.    The lattices we  used  are   displayed
Fig.~\ref{fig:latt}.

\subsection{Spectrum}

\begin{figure}
	\begin{center}
	\includegraphics[width=6cm]{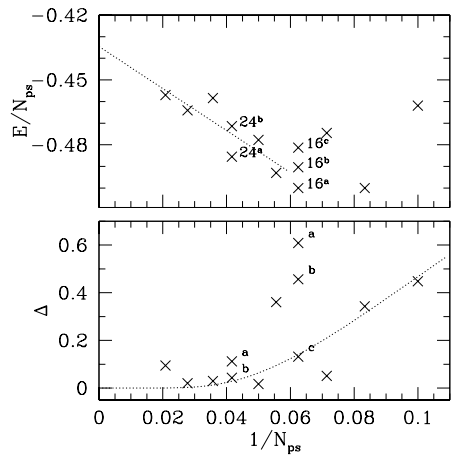}
	\end{center}

	\caption[99]{Top: Ground-state  energy  per site.   The dotted
	line   is     a    least-square fit    of     the   data   for
	$N_{ps}>16$.  Bottom:  energy   gap  between the  ground-state
	multiplet and  the first excited state.  The  dotted line is a
	guide to the eye  ($\sim e^{-aN_{ps}}$).  When different values are
	plotted for  the same  $N_{ps}$, they correspond   to different
	shapes   (see          $N_{ps}=16$       and      24        in
	Fig.~\ref{fig:latt}).}\label{fig:gap}

\end{figure}

\begin{figure}
	\begin{center}
	\includegraphics[width=6cm,bb=18 420 592 718]{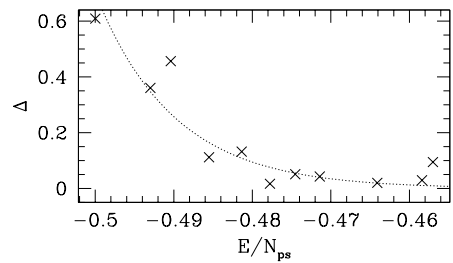}
	\end{center}

	\caption[99]{Gap as a    function  the energy  per   site  for
	$N_{ps}>12$ (data  of Fig.~\ref{fig:gap}). The  line is a guide to
	the eye.}\label{egap}

\end{figure}

The    ground-state   energy    per      hexagon   is   plotted     in
Fig.~\ref{fig:gap}. From this data  we can estimate that  $<\mu>\simeq
-0.44 \pm  0.02$  in the thermodynamic   limit.  It is  interesting to
compare  this   value with the   energy of  a simple   4-$\mu$ problem
$\mathcal{H}=\mu_1+\mu_2+\mu_3+\mu_4$      (with  periodic    boundary
condition so that  every site is neighbor  of the 3 others), which has
$<\mu>=-\frac{1}{2}$ in its  ground-state\footnote{It turns out  that,
as a  finite size effect, the $N_{ps}=12$  and 16 systems also exactly
have $<\mu>=-\frac{1}{2}$   in  their ground-state.}  The  energy  gap
between the ground-state multiplet and the first excited state is also
shown  in Fig.~\ref{fig:gap}. This quantity  probably goes  to zero in
the thermodynamic limit.  Fig.~\ref{egap} shows the  gap as a function
of the ground-state  energy per hexagon.   It appears that the samples
with  the largest gap  are those which   energy is significantly lower
that the thermodynamic estimate ($<\mu>\simeq -0.44  \pm 0.02$).  This
also points to a gapless spectrum in the limit of large systems.

The dispersion relation of  the first excited states usually  provides
some useful insight.   However, due to  the extensive degeneracy, this
brings no information for  $\mu$ model because the dispersion relation
can    be shown  to be perfectly    flat.  Let $\left|k\right>$ be  an
eigenstate  with   momentum $k$ and   energy  $E_k$.   The   new state
$\left|k+q\right>=\sum_h{e^{iq{\cdot}r_h}\mut(h)}\left|k\right>$  has,
by   construction, momentum $k+q$.    Because  the $\mut$ commute with
$\mathcal{H}_\mu$,  $\left|k+q\right>$  is still   an  eigenstate with
energy $E_k$. This  property is just  a  consequence of the  fact that
acting   with  a  $\mut$    creates  a  {\em    localized} zero-energy
excitations.

From  these data we propose  two possible scenarios:  1) An additional
ground-state degeneracy associated     to some spontaneous    symmetry
breaking in the thermodynamic limit.   As we will  explain, it may  be
that the ground-state orders in the three-sublattice pattern discussed
in section~\ref{sssec:abc}.  As  we  will explain, this scenario  does
not seem to be  the most likely.  In  particular, if the system was  a
three-sublattice crystal  in the $\mu$  variables,  the spectrum would
have  a ground-state with  a small\footnote{ 3  or 6 depending whether
the  axis symmetry  is   also spontaneously broken.}  quasi-degeneracy
(ignoring the exponential  factor  coming from the $\mut$  degrees  of
freedom) separated  by a {\em gap} to  higher excitations.   This does
not  seem to be  the case since we  could not identify  a small set of
energy levels  adjacent to the ground-state  above which a significant
gap could exist.  2) In the  second scenario the low-energy states may
correspond to a gapless mode of excitations in the system. Although we
have no   precise  theoretical  prediction  for  the  nature   of such
(critical) excitations in this 2D dimer model, gapless excitations are
reminiscent  of   the  one-dimensional  analog  (which   has fermionic
critical excitations at low energy see section~\ref{sec:1d} concerning
the    $\mu$-model on    the chain).  According    to the  single-mode
approximation discussed   in paragraph~\ref{ssec:sq0}  these   gapless
excitations would only exist at a finite value of the momentum.

\subsection{Correlations}

We looked for the possibility of long-ranged $\mu_i\mu_j$ correlations
in  the ground-state.  We define  a static  structure factor $S(k)$ in
the usual way:
\begin{eqnarray}
	S(k)&=&\frac{1}{N_{ps}} \left<0\right|
			\mu_k \mu_{-k} 
			\left|0\right> \\
	&=& \frac{1}{N_{ps}} \sum_{i,j} e^{-i k \cdot \left(r_i-r_j\right)}		
		\left<0\right| \mu_i\;\mu_j\left|0\right>
	\label{Sk}
\end{eqnarray}
where
\begin{equation}
	\mu_k=\sum_i e^{-i k \cdot r_i} \mu_i
\end{equation}
These  calculations were done numerically in  a  reduced Hilbert space
where $N_c\sim N_{ps}/2$ conserved quantities  (made out of $\mut$ and
$\tilde{T}$ operators)   are    fixed to be   $\pm1$   (non-degenerate
representation). By construction the spectrum does not depend on these
choices (that is the origin of the entropy) but it is also possible to
check that  $\left<\mu_i\;\mu_j\right>$   correlations do  not  depend
either on the  sector.   However we  stress  that it is in   principle
possible to have different correlations  in a ground-state which would
be  a   linear  combinations  of   the   ground-states   of  different
sectors. This  is similar to  the question of dimer-dimer correlations
discussed previously.   We have not  investigated these  effects which
are related to the possible ordering pattern which  may be selected by
small perturbations in the ground-state manifold.

\begin{figure}
	\begin{center}
	\includegraphics[width=4cm]{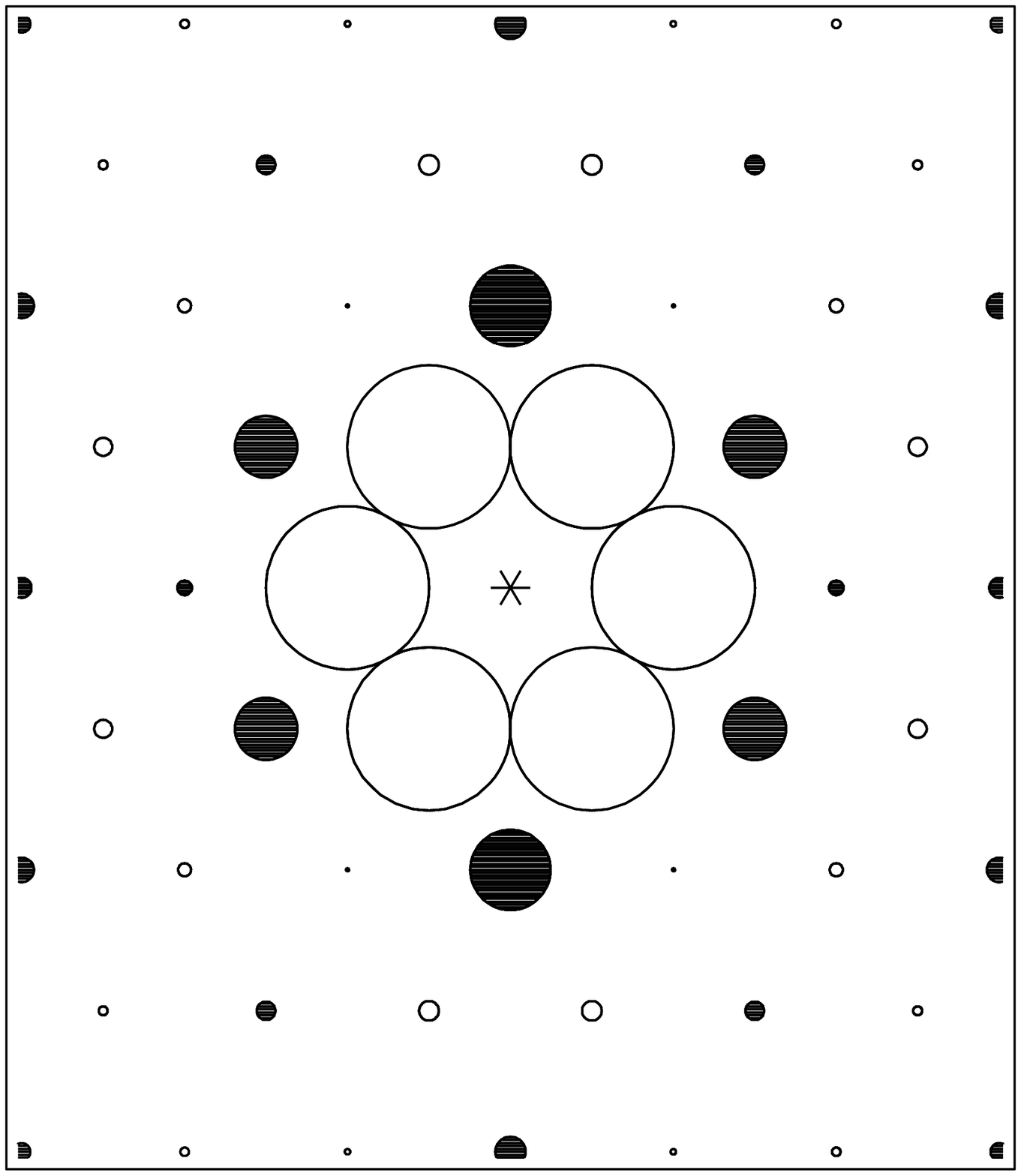}
	\end{center}

	\caption[99]{                                     Correlations
	$\left<\mu_0\mu_i\right>^c=\left<\mu_0\mu_i\right>-\left<\mu_0\right>^2$
	in a  48-hexagon system. The site  $0$ is at  the center.  The
	radius      of    the    circles     is    proportional     to
	$|\left<\mu_0\mu_i\right>^c|$.     Empty    circles   indicate
	negative  correlations and   the     black  ones  are      for
	$\left<\mu_0\mu_i\right>^c\ge0$ }\label{mumu48}

\end{figure}

\begin{figure}
	\begin{center}
	\includegraphics[width=5.5cm]{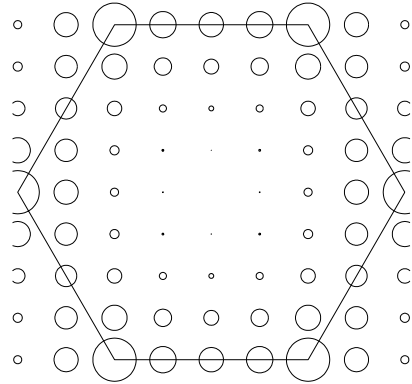}

	\end{center}              \caption[99]{Structure        factor
	$S(k)=\frac{1}{N}\left<\mu_{-k}\mu_k\right>$   represented  in
	the  first   Brillouin zone  of   the triangular   lattice for
	$N_{ps}=48$ (144 kagome sites).  The  radius of the circle  is
	proportional to $S(k)$.  $S(k)$ has  a trivial divergence   at
	$k=0$  which is due to the  fact that the operators $<\mu_i>$
	is  non-zero  at every   site.   This peak   at $k=0$ is   not
	represented here. }\label{sq}

\end{figure}

\begin{figure}
	\begin{center}
	\includegraphics[width=7cm]{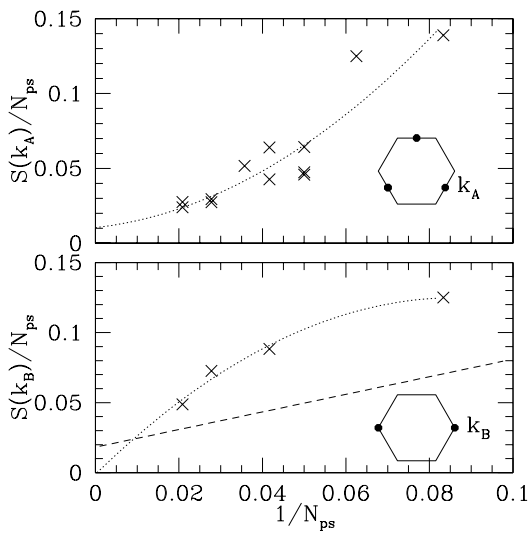}
	\end{center}

	\caption[99]{   Top:  structure  factor  at  $k=k_A$.  Bottom:
	Structure factor at $k=k_B$.  The dotted lines are obtained by
	a  least-square  fit of the    form $a+bN_{ps}^{-1}+cN_{ps}^{-2}$.   The
	dashed line is  the mean-field (Eq.~\ref{eq:Eabc})  prediction
	for $S(k_B)/N_{ps}$.    The  quick reduction of  the  crystal order
	parameter with the system  size (compared with  the mean-field
	result) suggests that the crystal is unstable to fluctuations.
	The  analysis  of  the related  susceptibilities confirms that
	$S(k)/N_{ps}$  is    indeed likely  to    extrapolate  to zero when
	$N_{ps}\to\infty$.}  \label{sab}

\end{figure}

The   results  are  summarized  in    Figs.~\ref{sq},  ~\ref{sq_}  and
\ref{sab}.  Fig.~\ref{sq}  clearly  indicate that the  most  important
correlations appear   at the  border  of  the Brillouin    zone.  More
precisely the corners of  the Brillouin zone $k_B=(\pm4\pi/3, 0)$ and
the     middle    points of  the  borders     of    the Brillouin zone
$k_{A_1}=(0,2\pi/\sqrt{3})$,      $k_{A_2}=(\pi,-\pi/\sqrt{3})$    and
$k_{A_3}=(\pi,\pi/\sqrt{3})$ are  the reciprocal lattice  points where
the correlations are the strongest. $k_B$ correspond to a 3-sublattice
structure  whereas  $    k_{A_i}$  is  related  to    a   2-sublattice
(stripe-like) order.  A (weak) tendency to a 3-sublattice ordering can
been  seen directly in  Fig.~\ref{mumu48} which represents real-space
correlations in the ground-state of the $N_{ps}=48$ sample (144 kagome
sites).   Almost  all  the sites   with a  positive correlation (black
circles) are  locate  on  the     same sublattice  (according to     a
3-sublattice decomposition) as the reference site.

To check  whether these correlations could  remain long-ranged  in the
thermodynamic  limit   we plot the  $S(k)/N_{ps}$    as a  function of
$N_{ps}$ (see Fig.~\ref{sab}).  As a   result, $S(k)/N_{ps}$ seems  to
extrapolate   to a very  small   (possibly 0) in  the  limit  of large
systems.   This suggest neither  2-  nor 3- sublattice ``crystalline''
order in the expectation values of  the $\mu$ operators.  However, the
data at $k=k_B$ should be compared with the mean-field state described
in section~\ref{sssec:abc}.  According to the expectation values given
by Eq.~\ref{eq:<muabc>} we  should have $S(k)/N_{ps}\simeq 0.0182$  in
the  thermodynamic limit.   While  the extrapolation  of the numerical
results of  Fig.~\ref{sab}   can not  distinguish  such  a small order
parameter from a disordered (or critical) phase, the prediction of the
mean-field approximation (dashed line  in Fig.~\ref{sab}) turns out to
be  significantly different from  the exact  ones.  In the  mean-field
approximation     $S(k_B)$   is     given  by     $S(k_B)/N_{ps}\simeq
0.0182+0.6279/N_{ps}$  (the $1/N_{ps}$ contribution  is just the local
contribution of  a given site  and its six  neighbors).   On the other
hand  the  exact value  of  $S(k_B)/N_{ps}$  decays  much faster  with
$N_{ps}$.  This  means  that   the reduction  of   the  crystal  order
parameter  $S(k_B)/N_{ps}$  with the  system size  is mainly caused by
long wave-length fluctuations rather than by local contributions. This
is a serious indication that the  three-sublattice crystal is unstable
with respect to these fluctuations.

\subsection{Static susceptibility}
To get  more insight on the  possibility of  some crystalline order in
the $\mu$ variables we calculated the static susceptibilities $\chi(k)$:
\begin{equation}
	\chi(k)=\frac{1}{2N_{ps}}\frac{\partial < \mu_k + \mu_{-k} >}
			{\partial \lambda}
\end{equation}
where  $\lambda$ is  the strength of  a infinitesimal symmetry-breaking
perturbation:
\begin{eqnarray}
	\mathcal{H}_\lambda &=& \sum_i \mu_i -\frac{1}{2}\lambda \left(\mu_k+\mu_{-k}\right)
	\label{Hl}
\end{eqnarray}
$\chi(k)$ is obtain numerically by measuring the expectation value
of $\mu_k+\mu_{-k}$ in the ground-state of the Hamiltonian Eq.~\ref{Hl}
in the presence of a small perturbation. The susceptibility is obtained
by extrapolating the result to $\lambda=0$.


\begin{figure}
	\begin{center}
	\includegraphics[width=6.5cm]{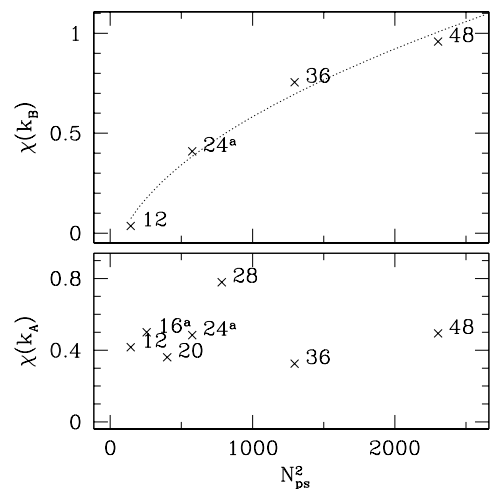}
	\end{center}

	\caption[99]{  Static  susceptibility $\chi(k)$  for the  wave
	vectors  ($k=k_A$   and $k=k_B$)   where correlations are  the
	strongest.   Long-range     order with  spontaneous   symmetry
	breaking would imply $\chi\sim N^2$,  which does not appear to
	be satisfied. The  dotted line is  a  least square fit of  the
	form $\chi(k_B)\simeq aN+b$.  }
	\label{chi}
\end{figure}

The static susceptibility  is a rather sensitive  probe since  it must
diverge  as $N_{ps}^2$  in    systems  that spontaneously  break   the
translation symmetry  in the  thermodynamic limit.\cite{santoro99}  On
the  other hand  it remains   finite if there   is no  ordering at the
corresponding wave-vector.  The   results for $k=k_A$  and $k=k_B$ are
displayed in Fig.~\ref{chi}. $\chi(k)$ shows no tendency to diverge at
$k=k_A$ and the increase with $N_{ps}$ of $\chi(k_B)$ is significantly
slower  than $\sim  N_{ps}^2$,  as suggested   by the rather  good fit
obtained    with   $\chi(k_B)\simeq  aN_{ps}+b$    (dotted   line   in
Fig.~\ref{chi}).  For these reasons we  think that the system does not
develop  long-ranged $\mu_i\mu_j$  correlations in  the  thermodynamic
limit.   The data   for   $\chi(k_B)$  (which neither  diverges   like
$N_{ps}^2$ not stay constant) might be interpreted as a proximity to a
{\em critical point} where a three-sublattice structure would appear.

\subsection{Long-wavelength fluctuations}
\label{ssec:sq0}

\begin{figure}
	\begin{center}
	\includegraphics[width=7cm]{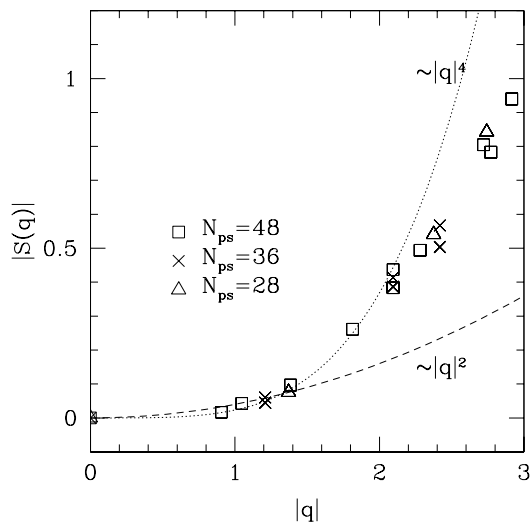}
	\end{center}
	\caption[99]{Long-wavelength behavior of the structure  factor
	$S(k)$ for  different sample sizes. }
	
	\label{sq_}	
\end{figure}

We now turn to the analysis of the long-wavelength fluctuations in the
system.  The structure factor  $S(k)$ is represented  as a function of
$|k|$ in Fig.~\ref{sq_}.  $S(k)$ seems to vanish as least as $S(q)\sim
|k|^2$ and  probably faster.  A  $\sim |k|^4$ behavior looks plausible
and  is reminiscent of  quantum Hall effect.\cite{prange} As explained
before, the dispersion  relation is flat  in this model.  However, one
may be interested in the excitations that can be created by the action
of   the $\mu$ operators  {\em  only} (excluding the  $\mut$).  Such a
variational excited state with momentum $k$  can be constructed in the
from the ground-state in spirit of the single mode approximation:
\begin{equation}
	\left| k \right> = \mu_k \left| 0 \right>
\end{equation}
The energy (relative to the ground-state) of $\left| k \right>$ is
\begin{equation}
	\omega(k)=
	\frac{\frac{1}{2N_{ps}}
	\left[\mu_{-k},\left[\mathcal{H},\mu_k \right] \right]}
	{S(k)}
\end{equation}
Since the  numerator  (oscillator  strength)   behaves, as  usual,  as
$\sim|k^2|$ at small  $|k|$, we find that $\left|  k \right>$ is not a
low energy  excitations  as soon  as $S(k)$  vanishes  like $|k|^2$ or
faster - which seems to be the case.  This is suggestive of a non-zero
gap for zero-momentum excitations.

\section{Discussion and conclusions}

We have introduced a  QDM on the kagome lattice  with a kinetic energy
which allows  from  3 to  6 dimers to   resonate around hexagons.  The
crucial difference  with  previous QDM    is   that dimer move    with
amplitudes which have non-trivial signs  inherited from the underlying
spin-$\frac{1}{2}$ model.   Exploiting the algebraic properties of the
conserved quantities ($\mut$ operators) we showed that the model as an
extensive entropy at zero temperature - $\frac{1}{6}\ln(2)$ per kagome
site - and is a dimer liquid.

The starting point of this study was the spin-$\frac{1}{2}$ Heisenberg
model on the kagome lattice.   Concerning this problem our main result
is that a  high density of singlet  states at low  energy might have a
real quantum origin and may  not just be  the  remainder of the  local
degeneracies of the classical model.  The mechanism which produces the
entropy of  the $\mu$ model  is new in the  sense that  one has to use
non-local degrees of  freedom in order  to compute  the spectrum in  a
representation which  eliminates the degeneracy.  We  are not aware of
any other model with a similar behavior.  Going from the KAFH model to
the $\mu$ model we made some  crude approximations.  The first one was
to reduce the spin Hilbert space to a short-range RVB one. We provided
several arguments to  support  this approximation but some  additional
studies would be required to analyze this question further. The second
drastic approximation was to reduce the  dimer dynamics induced by the
Heisenberg  interaction  to that   of  the $\mu$-model  and  its  {\em
signs}. This  can be  qualitatively   justified for  the  KAFH  in the
temperature regime corresponding  to  the low-temperature peak of  the
specific heat. If  that  picture is correct   the  degrees of  freedom
involved   in that peak would correspond   to  the $\sim N_{ps}/2=N/6$
uncoupled degrees of freedom of  the $\mu$-model.  The $\mu$-model can
be defined on any  lattice made of  corner-sharing triangles, it could
therefore provide a rather general explanation  for a large entropy at
low temperature  in  the corresponding  frustrated  spin-$\frac{1}{2}$
models.   Determining if some  order eventually develops at much lower
temperatures would amount  to analyze a degenerate perturbation theory
in the ground-state manifold of the $\mu$-model.

An important  question is to  know whether the  $\mu$-model realizes a
new phase or if it  is at a critical   point.  We have shown that  the
most  serious   candidate  for  an  ordered  phase,    if any, is  the
three-sublattice   crystal.  However   we   gave several   indications
(spectrum, correlations and  susceptibility) suggesting that it is not
stable.   Instead we suggest  that the system  might  be at a critical
point.  If we think of a dimer model  as a system of hard-core bosons,
it  is interesting to   compare our findings  with some  known bosonic
phases.   Let us first come  back to the  gapped RVB state realized in
the solvable QDM of Ref.~\onlinecite{msp02}.  That state, which is the
equal-amplitude     superposition   of     all  dimer   configurations
(Rokhsar-Kivelson~\cite{rk88}  state), is very  similar to a Bose {\em
condensate} in  the sense  that its wave-function  can be  obtained by
putting  all  dimers in the  same  zero-momentum state\footnote{If one
first  neglects  the hard core repulsion  of  dimers, one  can build a
state where all the dimers are in the same zero-momentum wave-function
$\phi(r)=1$.  The    many-dimer  wave      function  is of      course
$\psi(r_1,\cdots,r_n)=\phi(r_1)\cdots\phi(r_n)=1$             (already
symmetrized)   whatever      the   dimer   positions  $r_1,\cdots,r_n$
are.   Projecting this state (Gutzwiller   like) onto the  space of no
double-occupancy  precisely gives the equal-amplitude superposition of
all  fully-packed   dimer configurations.}.   The important difference
with a conventional superfluid  is that the  dimer model has no $U(1)$
gauge  symmetry (and therefore no  gapless ``sound'' mode or conserved
integer    charge)      but    a    discrete     $\mathbb{Z}_2$  gauge
symmetry~\cite{msp02}.  With that  comparison in mind the ground-state
of the $\mu$-model would neither be a  condensate nor a crystal but it
has {\em  gapless} excitations.   This is  rather unusual in  a  model
which has no continuous symmetry at all.  In addition  our model has a
structure factor $S(q)$  for  $\left<\mu\mu\right>$ correlations which
decays $|q|^2$   or faster in  the limit   $q\to0$.  In a  single-mode
approximation this would imply a gap for $q=0$ excitations.

In  order to   study  the spectrum   of   the $\mu$-model   we used  a
representation in which  the  degrees of freedom responsible  for  the
extensive  entropy are frozen.  The $\mu$-model  is local in the dimer
variables but the  effective Hamiltonian describing the non-degenerate
spectrum turned out  to be non-local  in terms of the original dimers.
These effective long-ranged interactions  between physical  degrees of
freedom  might be an important  ingredient  for the  appearance of  an
exotic  liquid.    From  this  point of   view   there might   be some
similarities between  the $\mu$-model and some two-dimensional quantum
systems of Bosons with long-ranged interactions.  Exotic liquid states
which  are not   superfluid  have  been  proposed for  these  systems,
including  a quantum  hexatic  phase~\cite{kklz91,fs97}.  On the other
hand  a critical Bose fluid  can exist without long-range interactions
and  such    a     phase    can     be    stabilized  by        cyclic
ring-exchanges.\cite{pbf02,bp02}  A   striking feature of   these  new
phases is the  existence of gapless excitations  along  {\em lines} in
the Brillouin  zone,   as in  a  Fermi  liquid.   We have  no   direct
indication of such a  behavior in the  kagome $\mu$-model, except  for
the existence of a fermionic  representation,  but such a scenario  is
certainly  an interesting possibility that should  be tested in future
studies.

{\em Acknowledgments}.---  We are  grateful to C.~Lhuillier,  F.~Mila,
R.~Moessner  and      M.~Oshikawa      for  several           fruitful
discussions. Numerical diagonalizations of QDM models were done on the
Compacq alpha server of the CEA under project 550.

\appendix

\section{Zeng and Elser's pseudospins representation of dimer coverings}
\label{sec:ZE}

Zeng and    Elser  (ZE)   realized~\cite{ez93,ze95}   that   a   close
correspondence between Ising  configurations of pseudospins sitting on
hexagons and  dimer configurations on  the  kagome lattice  (within  a
given topological sector) could  be used. We used  this representation
in a previous work~\cite{msp02} to  define an exactly solvable quantum
dimer model. In this section we review the pseudospin representation.

\subsection{$\sigma^z$ component}

We   first  need   an   (arbitrary)    reference dimer   configuration
$\left|D_0\right>$.    We   will   associate  a  pair   of  pseudospin
configurations   $\left\{\sigma^z_h=\pm 1\right\}_{h\in{\rm hex.}}$ to
any dimer configuration $\left|D\right>$ (belonging to the topological
sector of $\left|D_0\right>$) in the following way:

\begin{itemize}
	\item    Draw the  loops    of     the transition graph     of
	$\left<D_0|D\right>$.

	\item These loops must be considered as domain walls separating
	hexagons where the pseudospins  are up and hexagons where they
	are down.  This  can be done in  a consistent way  because $D$
	and $D_0$ are supposed to be in the same sector and any closed
	path will necessarily cross an even number of domain walls.

	\item There is a two-fold redundancy in the above prescription
	because up and down hexagons can be exchanged without changing
	the  loop  pattern.  Since there is  no  natural way to decide
	where is the interior  and where is the  exterior of  a closed
	loop   on  a   finite   sample,    a  pseudospin configuration
	$\left\{\sigma^z_h=\sigma(h)\right\}$    and     its  reversed
	counterpart $\left\{\sigma^z_h=-\sigma(h)\right\}$   represent
	the same dimer covering.
\end{itemize}

This establishes  a one-to-one correspondence between  dimer coverings
of  a  given topological sector and  {\em  pairs} of  pseudospin sates
related by a  global pseudospin  flip.  The proof can  be  done in two
steps:
\begin{itemize}
	\item If two dimer configurations $D$  and $D'$ are associated
	to  the  same  pseudospin  state (up   to a  global pseudospin
	reversal),    they   are  identical.  The   transition   graph
	$\left<D|D'\right>$ can  be  viewed  as   the   ``difference''
	between      graphs          $\left<D_0|D\right>$          and
	$\left<D_0|D'\right>$. This means that the $\left<D|D'\right>$
	will have  loops separating   regions where   the  pseudospins
	coincide in $D$ and $D'$ and  region where they are different.
	Since $D$ and $D'$ have the same  $\sigma^z$ on every hexagon,
	$\left<D|D'\right>$ cannot contain  any hexagon and  therefore
	contains no loop at all.

	\item Any  pseudospin configuration has  a corresponding dimer
	state.    The transition graph  between   $D_0$ and the  dimer
	configuration we are looking  for will  separate $\sigma^z=+1$
	hexagons from  $\sigma^z=-1$  ones.  The actual  path of these
	loops will depend  on  the  $D_0$  but for  a given  reference
	dimerization only one such path exists. The reason for this is
	easily  understood by looking  at  a single hexagon:  whatever
	$D_0$  maybe there is always  a single loop which surround this
	hexagon and this hexagon only.

\end{itemize}

We can check the above property by a direct counting.  On the one hand
we have $2^{N/3}$  pseudospins configurations and $2^{N/3-1}$ pairs of
non equivalent configurations. On the other hand there are $2^{N/3+1}$
(see Ref.~\onlinecite{elser89}) dimerizations on a kagome lattice with
$N$ sites and periodic boundary conditions.  The agreement is found by
remarking that  the number of dimerizations  has to be divided by {\em
four} to get the size of a  single topological sector.\footnote{It is a
property of the kagome lattice that  the four topological sectors have
exactly the same dimension.}

\subsection{Pseudo-spin flip operator $\sigma^x$}

One interest of ZE pseudospin representation  is that the $\sigma^x_h$
operator  which flips the pseudospin at  position $h$ can be expressed
in a simple way  in terms  of {\em  local} dimer operators.   It seems
that ZE  did   not realize    this  very  useful property   of   their
representation.   The   simplest  dimer   moves  involve  loops around
hexagons.  These  32  loops are represented  in Table~\ref{tab:loops}.
The corresponding operators
\begin{equation}
	\hat L_\alpha=\left|L_\alpha \right>\left< \bar{L}_\alpha\right|
	+
	\left| \bar{L}_\alpha\right>\left<L_\alpha\right|
\end{equation}
shift the dimers along   the loop $L^\alpha$  if  it is possible   and
annihilates the state otherwise.  We will now prove that $\sigma^x$ is
the sum of all the 32 kinetic operators of hexagon $h$:

\begin{equation}
	\sigma^x(h) = \sum_{\alpha=1}^{32} \hat L_\alpha
\end{equation}

The fact that this sum of dimer operators $\hat L^\alpha$ realizes the
spin  algebra is  not obvious.    In particular  the fact that   these
operators commute  at two  neighboring  hexagons $h$ and  $h'$ must be
verified since in general $\left[
\hat L_\alpha(h),\hat L_{\alpha'}(h')\right]\ne0$.

Consider an arbitrary dimerization $\left|D\right>$ in the vicinity of
an hexagon $h$.  The  crucial point is  that all the kinetic operators
$\hat L_\alpha(h)$ but one annihilate $\left|D\right>$.  This is a property
of  the kagome  lattice  that we already  used  before:  for any given
dimerization  one  and only  one loop  can surround   hexagon $h$.  So
$\left|D'\right>=\sigma^x(h)\left|D\right>$  is a dimer  configuration
which differs from $\left|D\right>$ by a single loop around $h$. Using
the $\sigma^z$ base  to represent dimer  coverings we know that such a
state  is  unique and is  the  state obtained from $\left|D\right>$ by
flipping   the  pseudospin  at in   $h$.   Thus   we  have shown  that
$\sigma^x(h)\sigma^z(h)=-\sigma^z(h)\sigma^x(h)$.

One can use a very similar reasoning  to show  that $\sigma^x(h)$ and  $\sigma^x(h')$
commute when $h{\ne}h'$ but  this result is most easily obtained
by the arrow representation.


\subsection{Counting dimer coverings with pseudospins}

The Ising  basis of ZE's  pseudospins  provides a  way of counting the
number $\mathcal{N}$  of dimer coverings on any  lattice  $K$ which is
the medial lattice of a trivalent one.  The result is
\begin{equation}
\mathcal{N}_{\rm dim. coverings}=\alpha 2^{N_{\rm ps}-1}
\label{eq:Ndcov_}
\end{equation}
where  $N_{\rm ps}$ is the  number of pseudospins  and $\alpha$ is the
number  of topological   sectors, it  is   related to  the   genus  by
$\alpha=2^{2g}$ in  the  two-dimensional cases.   The factor  $2^{-1}$
comes  from the fact that  a  dimer configuration  corresponds to  two
pseudospin  states.    Using  the Euler relation we   can   check  that
Eq.~\ref{eq:Ndcov_} indeed coincides with the result obtained with the
arrow representation (Eq.~\ref{eq:Ndcov3}):
\begin{equation}
	\mathcal{N}_{\rm dim. coverings}=2^{2g}2^{N/3+2-2g-1}=2^{N/3+1}
	\label{eq:Ndcov2}
\end{equation}


\section{$\mu$ kinetic operators and ZE pseudospins}
\label{sec:Sigmaz2Mu}

In order to write $\mu$ with $\sigma^z$ and $\sigma^x$ only we need to
express the sign $\epsilon(h)$ of Eqs.~\ref{eq:mu_def} and \ref{snout}
in terms  of the $\sigma^z$  operators on the neighboring hexagons. We
will do this with the help of the  arrow representation.  The argument
generalizes easily to the $\mut$ operators.

First draw  the  arrow representation  of  the reference  dimerization
$\left|D_0\right>$  in    the    vicinity  of hexagon    $h$,   as  in
Fig.~\ref{KagLoopWithArrows}.  Each neighboring hexagon of $h$ has two
arrows which  belongs to the  star of $h$.   These  arrows can either:
$A_1$)  point  toward the  exterior of $h$;   $A_2$) point  toward the
interior  of  $h$;  or $B$) point  in   two different directions.   In
Fig.~\ref{KagLoopWithArrows}, for example,  we have no hexagon in case
$A_1$, 2 in case $A_2$ and 4 in case $B$. First look  at the change in
$n_{\rm{out}}(h)$ when a single neighboring pseudospin is flipped with
respect to the reference configuration  (it has $\sigma^z=-1$). If the
corresponding  hexagon    is     of  type   $A_1$   (resp.     $A_2$),
$n_{\rm{out}}(h)$ is  decreased (resp.  increased) by  $2$  units.  If
that  hexagon is of type  $B$,  $n_{\rm{out}}(h)$ is  unchanged.  On a
state where a single pseudospin is down we have therefore showed that:
$\epsilon(h)=\epsilon_{\rm{ref}}(h)\prod_{i{\in}A_{\rm{ref}}(h)}\sigma^z(i)$
where $A_{\rm{ref}}(h)$ is  the set of the neighbors  of $h$ which are
of type $A_1$ or $A_2$ in the  reference dimerization.  $\epsilon_{\rm
ref}(h)$ is the value of $\epsilon(h)$ in the reference state.

Now look at  the value of  $\epsilon(h)$ when the neighboring pseudospins are
in  an arbitrary state  $\sigma^z(h_i)=\pm1$.   If two pseudospins are
flipped on hexagons which are themselves first neighbors, one arrow is
flipped twice and therefore remains  unchanged.  This adds to $\epsilon(h)$ a
$-1$ factor,  which multiplies the  single pseudospin factor discussed
above.       This    can  be    seen       on     the   example     of
Fig.~\ref{KagLoopWithArrows}.   If  the   pseudospin  of  the  hexagon
containing  the sites  $(2',3,3')$ (type  $A_2$)  is flipped the  sign
$\epsilon(h)$ changes but if hexagon $(1',2,2')$ (type $B$) is flipped $\epsilon(h)$
is not affected.   However, if {\em both}  hexagons are flipped $\epsilon(h)$
is unchanged too.  In  addition to the $\sigma^z(h_i)$ factors  coming
from hexagons belonging  to   $A_{\rm{ref}}(h)$  we must add   a  $-1$
contribution for  each {\em pair of  consecutive pseudospins which are
simultaneously in a $\sigma^z(h_i)=-1$ state}.  Such a factor is given
by:
\begin{equation}
	\prod_{i=1}^6 f\left[\sigma^z(h_i),\sigma^z(h_{i+1})\right]
\end{equation}
where $f$ is equal to   $-1$ when the two   pseudospins are both  down
$\sigma^z(h_i)=\sigma^z(h_{i+1})=-1$ and   $=1$ otherwise. The product
runs over   the six neighbors  $h_i$   of $h$ numerated  in  a  cyclic
way. $f$ can be explicitly written as a polynomial:
\begin{equation}
	f\left[\sigma^z_1,\sigma^z_2\right]
	=\frac{1}{2}\left(1+\sigma^z_1+\sigma^z_2 -\sigma^z_1\sigma^z_2\right)
\end{equation}
We eventually  have  an expression for $\epsilon(h)$  as  a polynomial of  ZE
pseudospins operators located on neighbors of $h$:
\begin{equation}
	\epsilon(h)=\epsilon_{\rm ref}(h)
		\prod_{i\in A{\rm{ref}}(h)}\sigma^z(i)
		\prod_{i=1}^6 f\left[\sigma^z(h_i),\sigma^z(h_{i+1})\right]
\label{eq:Sigmaz2Mu}
\end{equation}
The left-hand    side   is both   a  local   and reference-independent
operator. On the right-hand side, the information on the arrows in the
reference   configuration   is     present  at  several   places:   in
$\epsilon_{\rm{ref}}(h)$, in  the  set  of sites  $A_{\rm{ref}}(h)$ as
well as in the $\sigma^z$ operators.  It is possible to check directly
on this expression that the $\epsilon(h)\sigma^x(h)$ satisfy the $\mu$
algebra.


\section{Removing the extensive degeneracy}
\label{sec:red}

To perform  the explicit diagonalization, it is  useful to  remove the
extensive degeneracy.  Let us  denote by $C_\alpha$,  $\alpha=1,\ldots
N_c$,   the   commuting   and  independent  operators   introduced  in
Sec.~\ref{sssec:nodegrep}, and $c_\alpha=\pm  1$ their eigenvalues. We
can  decompose the Hilbert  space in eigenspaces $\{c_\alpha\}$, using
the projectors  $$P_\pm^\alpha=\frac{1}{2}  (1\pm C_\alpha)\;.$$ It is
then sufficient to work within the reduced  space where $c_\alpha$ are
all equal to $1$, for example. The reduced space  is then generated as
follows.  Consider the state $\left|\{s(h)=1\} ;\{I_k\} \right>$ where
all $s(h)=1$, and the integrals of  motion have some values $\{I_k\}$.
This state exists, is  unique and it has  the property of transforming
the  action $\mu(h)$ into that   of $\tilde \mu(h)$  for  all $h$.  We
project this state  to the eigenspace $\{c_\alpha\}=1$, $$\left|\Omega
;\{I_k\}\right>=\prod_\alpha P_+^\alpha
\left|\{s(h)=1\}  ;\{I_k\}\right>\;$$  and   then generate  the  whole
subspace  $\{c_\alpha=1\}$ by the  action  of the  monomials  in $\mu$
$$\prod_{h=1}^{N_{ps}} \mu(h)^{n_h}\left|\Omega ; \{I_k\} \right>\;,$$
with $n_h=0,1$.     Roughly  half of  the   spins can   be  eliminated
recursively  by the following procedure.    For spins belonging to the
sublattice $A$, we  bring $\mu(h)$   at  right using the   commutation
relations,   we transform    it   in  $\mut(h)$,    which    is  1  by
construction. Some of the spins can be  eliminated using the integrals
of motion $I_k$.  And finally, one of  the spins entering to a bow-tie
operator can be  eliminated by replacing  it with  the product  of the
other three spins  in the bow-tie and  the operator $T(h)$. The latter
operator is brought to  the right, transformed in $\tilde{T}(h)$ whose
value is 1. Similarly, one can construct the eigenspace for any values
of $\{c_\alpha\}$. The passage  from a sector of fixed  $\{c_\alpha\}$
to another  can be realized by  operators $\tilde \mu(h)$ belonging to
one  of the sublattices $B,C,D$.   Since  these operators commute with
the Hamiltonian, the spectrum of $\mathcal{H}_\mu$ will be the same in
all the sectors, which   proves  that the   global degeneracy of   the
spectrum is $2^{N_c}$.

The action of the  Hamiltonian on the reduced  basis is computed using
the  same procedure as above. The  fact that we  work in  a space with
dimension  divided by  $2^{N_{ps}/2}$ allows  us  to perform numerical
diagonalization for relatively large systems, up to $N_{ps}=48$.


\end{document}